\documentclass[english,aip,letter,superscriptaddress,floatfix,jcp,11pt]{revtex4}
\usepackage[T1]{fontenc}
\usepackage[latin9]{inputenc}
\usepackage{color}
\usepackage{babel}
\usepackage{verbatim}
\usepackage{float}
\usepackage{mathtools}
\usepackage{amsmath}
\usepackage{amssymb}
\usepackage{graphicx}
\usepackage[unicode=true,pdfusetitle,
 bookmarks=true,bookmarksnumbered=false,bookmarksopen=false,
 breaklinks=false,pdfborder={0 0 1},backref=false,colorlinks=false]
 {hyperref}

\makeatletter

\floatstyle{ruled}
\newfloat{algorithm}{tbp}{loa}
\providecommand{\algorithmname}{Algorithm}
\floatname{algorithm}{\protect\algorithmname}

\@ifundefined{textcolor}{}
{%
 \definecolor{BLACK}{gray}{0}
 \definecolor{WHITE}{gray}{1}
 \definecolor{RED}{rgb}{1,0,0}
 \definecolor{GREEN}{rgb}{0,1,0}
 \definecolor{BLUE}{rgb}{0,0,1}
 \definecolor{CYAN}{cmyk}{1,0,0,0}
 \definecolor{MAGENTA}{cmyk}{0,1,0,0}
 \definecolor{YELLOW}{cmyk}{0,0,1,0}
}

\usepackage{ae,aecompl}

\usepackage{amsthm,mathtools}
\theoremstyle{plain}

\numberwithin{thm}{subsection}

\numberwithin{lem}{subsection}

\usepackage{algorithm}
\usepackage[noend]{algpseudocode}

\makeatother

\begin{document}

\title{Large Scale Brownian Dynamics of Confined Suspensions of Rigid Particles}

\author{Brennan Sprinkle}

\affiliation{McCormick School of Engineering, Northwestern University, Evanston,
IL 60208}

\author{Florencio Balboa Usabiaga}

\affiliation{Courant Institute of Mathematical Sciences, New York University,
New York, NY 10012}

\affiliation{Center for Computational Biology, Flatiron Institute, Simons Foundation,
New York 10010, USA}

\author{Neelesh A. Patankar}

\affiliation{McCormick School of Engineering, Northwestern University, Evanston,
IL 60208}

\author{Aleksandar Donev}
\email{donev@courant.nyu.edu}

\affiliation{Courant Institute of Mathematical Sciences, New York University,
New York, NY 10012}
\begin{abstract}
\textcolor{black}{We introduce methods for large scale Brownian Dynamics
(BD) simulation of many rigid particles of arbitrary shape suspended
in a fluctuating fluid. Our method adds Brownian motion to the rigid
multiblob method {[}F. Balboa Usabiaga }\textcolor{black}{\emph{et
al.}}\textcolor{black}{, Communications in Applied Mathematics and
Computational Science, 11(2):217-296, 2016{]} at a cost comparable
to the cost of deterministic simulations. We demonstrate that we can
efficiently generate deterministic and random displacements for many
particles using preconditioned Krylov iterative methods, if kernel
methods to efficiently compute the action of the Rotne-Prager-Yamakawa
(RPY) mobility matrix and it ``square'' root are available for the
given boundary conditions. These kernel operations can be computed
with near linear scaling for periodic domains using the Positively
Split Ewald method. Here we study particles partially confined by
gravity above a no-slip bottom wall using a graphical processing unit
(GPU) implementation of the mobility matrix vector product, combined
with a preconditioned Lanczos iteration for generating Brownian displacements.
We address a major challenge in large-scale BD simulations, capturing
the stochastic drift term that arises because of the configuration-dependent
mobility. Unlike the widely-used Fixman midpoint scheme, our methods
utilize random finite differences and do not require the solution
of resistance problems or the computation of the action of the inverse
square root of the RPY mobility matrix. We construct two temporal
schemes which are viable for large scale simulations, an Euler-Maruyama
traction scheme and a Trapezoidal Slip scheme, which minimize the
number of mobility solves per time step while capturing the required
stochastic drift terms. We validate and compare these schemes numerically
by modeling suspensions of boomerang shaped particles sedimented near
a bottom wall. Using the trapezoidal scheme, we investigate the steady-state
active motion in a dense suspensions of confined microrollers, whose
height above the wall is set by a combination of thermal noise and
active flows. We find the existence of two populations of active particles,
slower ones closer to the bottom and faster ones above them, and demonstrate
that our method provides quantitative accuracy even with relatively
coarse resolutions of the particle geometry.}
\end{abstract}
\maketitle
\global\long\def\V#1{\boldsymbol{#1}}
\global\long\def\M#1{\boldsymbol{#1}}
\global\long\def\Set#1{\mathbb{#1}}

\global\long\def\D#1{\Delta#1}
\global\long\def\d#1{\delta#1}

\global\long\def\norm#1{\left\Vert #1\right\Vert }
\global\long\def\abs#1{\left|#1\right|}

\global\long\def\grad{\M{\nabla}}
\global\long\def\avv#1{\langle#1\rangle}
\global\long\def\av#1{\left\langle #1\right\rangle }

\global\long\def\P{\mathcal{P}}

\global\long\def\ki{k}
\global\long\def\wi{\omega}

\global\long\def\slip{\breve{\V u}}

\global\long\def\bu{\V u}
 \global\long\def\bv{\V v}
 \global\long\def\br{\V r}

\global\long\def\sM#1{\M{\mathcal{#1}}}
\global\long\def\Mob{\sM M}
\global\long\def\J{\sM J}
\global\long\def\S{\sM S}
\global\long\def\L{\sM L}

\global\long\def\N{\sM N}
\global\long\def\K{\sM K}
\global\long\def\slipN{\breve{\N}}

\global\long\def\aN{\widetilde{\N}}
\global\long\def\aK{\widetilde{\K}}
\global\long\def\aMob{\widetilde{\Mob}}

\global\long\def\epsN{\overline{\N}}

\global\long\def\slipW{\breve{\V W}}
\global\long\def\rot{\M{\Psi}}
\global\long\def\Rot{\M{\Xi}}

\global\long\def\eqd{\overset{d}{=}}

\section{Introduction}

The study of Brownian suspensions of passive and active particles
has become ubiquitous, particularly in the biological and physical
sciences \cite{BrownianDynamics_DNA,BrownianReview,Nanomotors_Kapral}.
Thermal fluctuations play an integral role in determining the dynamics
of active suspensions, for example, they affect the development of
a recently-discovered fingering instability \cite{Rollers_NaturePhys}
in suspensions of microrollers partially confined by gravity above
a no-slip wall bottom wall \cite{MagneticRollers}. Efficient simulations
of these systems, which correctly capture the effects of Brownian
motion, are essential for designing and understanding experimental
results as well as discovering new collective phenomena. The purpose
of this work is to design scalable and accurate numerical methods
for Brownian Dynamics (BD) simulation of suspensions of many passive
and/or active rigid particles of arbitrary shape. By scalable, we
mean that the computational complexity of the method should scale
(nearly) \emph{linearly} with the number of particles considered.
By accurate, here we mean that the temporal integrators used should
be second-order accurate in the deterministic setting (i.e., without
Brownian motion), and maximize the weak accuracy for a given computational
cost per unit simulation time. To our knowledge, this is the first
time either of these goals have been accomplished with \emph{controlled}
accuracy for a suspension of particles of arbitrary shape. Existing
methods based on uncontrolled multipole truncations \cite{StokesianDynamics_Brownian,FluctuatingFCM_DC,BoundaryIntegralWall_Adhikari}
are focused on spheres and spheroids and are difficult to generalize
to general particle shapes, though some special shapes like thin rods
have been tackled using slender body theory \cite{BD_Rods_Shaqfeh}.

In \cite{RigidMultiblobs} some of us proposed a flexible and scalable
\emph{rigid multiblob} method for simulation of many rigid bodies
(not necessarily spherical) in Stokes flow, in the absence of Brownian
motion. In the rigid multiblob method \cite{RigidMultiblobs,BrownianMultiBlobs},
complex particle shapes are constructed with tunable resolution (accuracy)
as a rigidly-connected cluster of spherical ``blobs''. In \cite{RigidMultiblobs}
some of us developed efficient preconditioned Krylov methods for solving
mobility problems in computational cost that scales (nearly) linearly
with the number of blobs used to construct the rigid particles. Here
we focus on efficient computation of stochastic displacements consistent
with the overdamped Langevin equation for non-spherical particles
proposed by some of us in \cite{BrownianMultiBlobs}. In the prior
work \cite{BrownianMultiBlobs} we assumed a small number of particles
and performed direct (Cholesky) factorization of mobility matrices
to compute Brownian increments, which scales as the cube of the number
of blobs and is infeasible for many-body suspensions. In this work
we develop linear-scaling preconditioned iterative methods for computing
the stochastic increments of particle positions and orientations.
A second nontrivial challenge we address is the construction of consistent
and accurate temporal integrators. The widely-used Fixman midpoint
temporal integrator, generalized to include particle orientations
in \cite{BrownianMultiBlobs}, requires solving resistance problems,
which cannot be done in linear time with present methods \cite{RigidMultiblobs}.
Here we construct two temporal integrators that correctly capture
stochastic drift terms proportional to the divergence of the mobility
matrix, and require only solving mobility problems. While here we
only test these novel schemes with the rigid multiblob method \cite{RigidMultiblobs},
it is important to note that the same temporal integrators apply to
highly-accurate boundary integral formulations \footnote{As explained in detail in Appendix A of \cite{RigidMultiblobs}, the
rigid multiblob method can be seen as a first-kind boundary integral
method regularized in a physically-consistent way so as to ensure
the symmetry and positive definiteness of the mobility matrix, as
required for adding Brownian motion.} for Stokesian suspensions \cite{BoundaryIntegral_Pozrikidis,BoundaryIntegral_Periodic3D,BoundaryIntegral_SpheroidQBX}.
Furthermore, while we focus here on suspensions confined above a no-slip
wall, the methods we present here are rather general and can be applied
to other systems such as bulk passive or active suspensions.

In section \ref{sec:Nhalf}, we develop a scalable method to generate
the Brownian increments for the particles from the Brownian increments
of the individual blobs, which can themselves be computed using a
preconditioned Lanczos method \cite{SquareRootPreconditioning}, as
previously described for particles above a no-slip wall in \cite{MagneticRollers},
and for periodic suspensions in \cite{SpectralRPY}. In section \ref{Traction}
we propose a novel modification of the Euler-Maruyama (EM) scheme,
which involves solving only a \emph{single} additional mobility problem
in order to capture the Ito stochastic (thermal) drift required to
maintain the Gibbs-Boltzmann distribution at equilibrium. This is
a notable improvement over the EM method proposed in \cite{BrownianMultiBlobs}
which requires \emph{two} additional mobility solves to compute the
drift using a \emph{random finite difference} (RFD). In section \ref{sec:Slip}
we propose a novel trapezoidal scheme which also captures the correct
thermal drift by solving only a single additional mobility problem,
and is second order accurate in time for deterministic calculations.
While the scheme is formally only first-order weakly accurate in the
stochastic setting, the improved deterministic accuracy translates
to substantially improved weak accuracy, as we demonstrate numerically.

In sections \ref{sec:TwoBooms} and \ref{sec:ManyBooms}, we validate
the new temporal integrators and compare their efficiency/accuracy
tradeoffs by examining equilibrium statistics for suspensions of passive
colloidal boomerangs confined above a wall. In section \ref{sec:BrownRoll}
we revisit some experimental and computational investigations done
in \cite{MagneticRollers} for dense uniform suspensions of rotating
colloids (microrollers) above a planar wall. In this prior work \cite{MagneticRollers},
a large mismatch was observed between experimental measurements of
the steady-state mean suspension velocity and estimates based on the
minimally-resolved Brownian dynamics computations \cite{MagneticRollers}.
Here we are able to simulate a dense uniform suspension of microrollers
with much higher resolution. The higher resolution allows us to better
resolve the hydrodynamic interactions between the particles and make
quantitative predictions that are sufficiently accurate to be directly
compared to experiments.

\section{\label{sec:BD} Brownian Dynamics for Rigid Bodies}

We consider a suspension of $N_{b}$ passive or active rigid bodies
(particles) suspended in a fluctuating Stokesian fluid. For body $p\in[1,\ldots,N_{b}]$,
we will follow a reference\emph{ tracking point} with Cartesian position,
$\V q_{p}\left(t\right)$. The orientation of body $p$ relative to
the tracking point will be denoted by $\V{\theta}_{p}\left(t\right)$.
For simplicity and increased generality, we make the bulk of the discussion
in this work agnostic to the choice of coordinates for $\V{\theta}_{p}$
and assume that the representation is a scalar in two dimensions or
a three-dimensional vector in three dimensions. In practice, however,
we use unit quaternions in three dimensions, as discussed in detail
in \cite{BrownianMultiBlobs}. The unit norm constraint of the quaternion
can be handled simply by updating orientation using quaternion multiplication
(rotations) instead of addition, as detailed in Appendix \ref{Add:Quat}.
We denote the generalized position of body $p$ as $\V Q_{p}\left(t\right)=\left[\V q_{p}\left(t\right),\V{\theta}_{p}\left(t\right)\right]$
and denote the many-body configuration with $\V Q=\left[\V Q_{p}\right]$.
To each body $p$, we prescribe an applied force $\V f_{p}$, and
an applied torque $\V{\tau}_{p}$, and denote the generalized force
on body $p$ with $\V F_{p}=\left[\V f_{p},\V{\tau}_{p}\right]$ and
write $\V F=\left[\V F_{p}\right]$. The prescription of external
(non-conservative) forces and torques is one way in which we may model
active bodies, the other, active slip, is discussed in more detail
in \cite{RigidMultiblobs} and summarized in section \ref{sec:multN}. 

Given forces and torques, our aim is to find the rigid body velocities
$\V U=\left[\V U_{p}\right]$, where the generalized velocity $\V U_{p}=\left[\V u_{p},\V{\omega}_{p}\right]$
is composed of a translational velocity $\V u_{p}$ and a rotational
(angular) velocity $\V{\omega}_{p}$. A central object in the overdamped
Langevin equations for the suspension is the configuration-dependent
\textit{body mobility matrix} $\N\left(\V Q\right)$. In a deterministic
setting, the symmetric positive-definite (SPD) matrix $\N$, relates
the generalized velocities with the generalized forces, $\V U=\N\V F.$
The application of the body mobility matrix, i.e., the computation
of $\V U=\N\V F$ given $\V F$, is referred to as the \textit{mobility
problem.} Its inverse problem, the \textit{resistance problem}, involves
finding the forces and torques given prescribed rigid-body motions,
i.e., computing $\V F=\N^{-1}\V U$. By combining the rigid multiblob
method with preconditioned iterative solvers, one can solve a mobility
problem efficiently in linear time, however, the solution of resistance
problems is much more expensive and does not scale linearly \cite{RigidMultiblobs}.

For a suspension of rigid bodies, the configuration evolves according
to the overdamped Langevin Ito BD equation, 
\begin{equation}
\frac{d\V Q}{dt}=\V U=\N\V F+k_{B}T\left(\partial_{\V Q}\cdot\N\right)+\sqrt{2k_{B}T}\ \N^{1/2}\sM W,\label{Langevin}
\end{equation}
where $\sM W$ is a collection of independent white noise processes
\cite{BrownianMultiBlobs}. Here the ``square root'' of the mobility
$\N^{1/2}$ is any matrix, \emph{not} necessarily square, that satisfies
the fluctuation-dissipation relation $\N=\N^{1/2}\left(\N^{1/2}\right)^{T}$.
We will refer to the term $k_{B}T\ \partial_{\V Q}\cdot\N$ as the
\emph{stochastic} or \textit{thermal drift} (or sometimes simply the
\textit{drift}) since it has its origin in the stochastic interpretation
of the noise; this drift term would disappear if the so-called kinetic
or Klimontovich interpretation of the noise is used \cite{KineticStochasticIntegral_Ottinger}.
Efficient generation of this term will be the most challenging part
of this work and is discussed in detail in section \ref{sec:RFD}.
As a prelude, in subsection \ref{sec:multN} we briefly review the
methods proposed in \cite{RigidMultiblobs} to efficiently compute
the deterministic displacements, $\N\V F$. Then, in subsection \ref{sec:Nhalf},
we propose a scalable iterative method for computing the Brownian
displacements over a time interval $\D t$, $\sqrt{2k_{B}T\,\D t}\ \N^{1/2}\V W$,
where $\V W$ is a vector of independent standard Gaussian random
variables.

\subsection{\label{sec:multN}Solving Mobility Problems}

We discretize the rigid bodies using a rigid multi-blob model, wherein
rigid bodies are treated as rigid conglomerations of beads, or ``blobs'',
of hydrodynamic radius $a$. The blobs comprising a given rigid body
$\mathcal{B}_{p}$ have positions $\V r^{(p)}=\left[\V r_{i}\ |\ i\in\mathcal{B}_{p}\right]$.
Given a rigid body velocity $\V U_{p}=\left[\V u_{p},\V{\omega}_{p}\right]$,
the \textit{geometric }\textit{\emph{block matrix}} $\K$ that converts
rigid body motion into blob motion is defined as \cite{StokesianDynamics_Rigid}
\begin{equation}
\left(\K\V U\right)_{i}=\V u_{p}+\V{\omega}_{p}\times\left(\V r_{i}-\V q_{p}\right),\hspace{0.5cm}p\in1,\ldots,N_{B}\text{ and }i\in\mathcal{B}_{p}.\label{K}
\end{equation}
Using $\K$, a slip condition on the rigid bodies can be compactly
expressed as 
\begin{equation}
\dot{\V r}=\K\V U-\slip,
\end{equation}
where $\slip$ is a prescribed slip velocity of the fluid at the locations
of the blobs. Physically, $\slip$ could account for an active boundary
layer \cite{BoundaryIntegralGalerkin,BoundaryIntegralWall_Adhikari}.
However, in this work, we will find a great deal of utility in prescribing
$\slip$ in such a way as to help generate the stochastic terms in
equation (\ref{Langevin}). The force and torque balance conditions
on the particles can be expressed using the adjoint of $\K$, $\K^{T}\V{\lambda}=\V F$
\cite{StokesianDynamics_Rigid}.

The hydrodynamic interactions between blob $i$ and $j$ are captured
by the $3\times3$ mobility matrix $\M M_{ij}$, which gives the velocity
$\dot{\V r}_{i}$ of blob $i$ given a force $\V{\lambda}_{j}$ on
blob $j$, $\dot{\V r_{i}}=\M M_{ij}\V{\lambda}_{j}$. The symmetric,
positive semi-definite matrix $\Mob$ composed of the blocks $\M M_{ij}$
is termed the \textit{blob-blob mobility matrix}. The construction
of $\Mob$ for a rigid multiblob must account for the finite hydrodynamic
radius of the blobs, $a$, as well as the geometry of the domain.
In the case of a three dimensional unbounded domain, the well-known
Rotne-Prager-Yamakawa (RPY) tensor \cite{RotnePrager,RPY_Shear_Wall}
can be used to construct $\M M_{ij}$, and the action of $\Mob$ on
a vector can be computed in linear time using a fast multipole method
\cite{RPY_FMM}. For periodic domains, we can use the Positively Split
Ewald (PSE) method \cite{SpectralRPY} to compute the action of the
RPY-based mobility $\Mob$ on a vector. A generalization of the RPY
kernel to particles confined above a single no-slip wall, the Rotne-Prager-Blake
tensor, is given in \cite{StokesianDynamics_Wall} and we will use
it in section \ref{sec:Results}. For general fully-confined domains,
an on-the-fly procedure to calculate $\Mob$ has been proposed in
\cite{BrownianBlobs,RigidIBM}. Note that the action of $\Mob$ can
be interpreted as a physically-regularized single-layer (first-kind)
boundary integral operator (see appendix A of \cite{RigidMultiblobs}).

Given $\Mob$, we can write the mobility problem as a linear system
\begin{align}
\Mob\V{\lambda} & =\K\V U-\slip\label{eq:mobU}\\
\K^{T}\V{\lambda} & =\V F,\label{align:FTBAL}
\end{align}
which can be written as the \emph{saddle-point} linear system, 
\begin{equation}
\begin{bmatrix}\Mob & -\K\\
-\K^{T} & \V 0
\end{bmatrix}\begin{bmatrix}{\V{\lambda}}\\
\V U
\end{bmatrix}=\begin{bmatrix}-\slip\\
-\V F
\end{bmatrix}.\label{eq:saddle}
\end{equation}
Using Schur complements, we can compactly write the solution to (\ref{eq:saddle})
as 
\begin{equation}
\V U=\N\V F+\N\K^{T}\Mob^{-1}\slip,\label{eq:sysDense}
\end{equation}
where we have identified the \emph{body mobility matrix} 
\begin{equation}
\N=\left(\K^{T}\Mob^{-1}\K\right)^{-1}.\label{eq:N_def}
\end{equation}
We note that exactly the same saddle-point system, with a mobility
matrix $\Mob$ computed using singular quadratures instead of the
RPY kernel, appears in a recently-developed first-kind Fluctuating
Boundary Integral Method (FBIM) for suspensions \cite{FBIM}.

In the case of many bodies, computing (the action of) $\N$ directly
from equation (\ref{eq:N_def}) is very inefficient if at all feasible.
In practice, we will solve mobility problems by solving (\ref{eq:saddle})
directly. Efficient, preconditioned Krylov solvers to solve this system
were developed in \cite{RigidMultiblobs}. The efficiency of these
solvers is dependent, primarily, on the speed at which the matrix,
$\Mob$, can be applied to a vector. If a linear-scaling method such
as a fast-multipole-method (FMM) \cite{RPY_FMM,OseenBlake_FMM} or
the PSE method \cite{SpectralRPY} are used, these methods will scale
near linearly (to within logarithmic factors) with the total number
of blobs. Following \cite{RigidMultiblobs}, here we will use direct
dense matrix-vector products implemented on a GPU to apply the Rotne-Prager-Blake
mobility. While this in principle scales quadratically with the total
number of blobs, modern GPUs are typically powerful enough for a direct
implementation of a matrix-vector product to outperform more sophisticated
techniques up to a fairly large number (hundreds of thousands) of
blobs \cite{MagneticRollers}. No matter how fast the matrix-vector
products with $\N$ (or equivalently $\Mob$) can be computed, solving
the system (\ref{eq:saddle}) is one of two bottlenecks in designing
efficient integrators to solve (\ref{Langevin}). We discuss the other
bottleneck next.

\subsection{\label{sec:Nhalf}Computing Brownian increments}

As mentioned in section \ref{sec:multN}, direct computation of $\N=\left(\K^{T}\Mob^{-1}\K\right)^{-1}$
is computationally infeasible for many bodies due to the dense matrix
inversions required. Direct computation of $\N^{1/2}$, therefore,
is still less practical in these situations. Our key insight to overcome
this is that $\N^{1/2}$ is not unique and doesn't need to be a square
matrix, it only needs to satisfy $\N=\N^{1/2}\left(\N^{1/2}\right)^{T}$.
This gives great freedom in choosing $\N^{1/2}$ so that its action
can be computed in linear time. We will assume here that we were able
to efficiently compute Brownian displacements for the individual blobs,
i.e., to compute $\Mob^{1/2}\V W$, where $\V W$ is a vector of independent
standard Gaussian random variables. This can be done using preconditioned
iterative methods for bodies near a no-slip wall \cite{MagneticRollers},
using the PSE method \cite{SpectralRPY} for periodic suspensions,
or using the FBIM \cite{FBIM} for fully confined or periodic suspensions.

Let us impose the random slip velocity $\slip=\sqrt{2k_{B}T/\D t}\ \Mob^{1/2}\V W$
in (\ref{eq:saddle}) \footnote{The random slip velocity $\slip$ can be derived by using fluctuating
hydrodynamics. For example, one can start with the coupled fluid+particle
equations given in Eq. (20) in \cite{RigidMultiblobs}, and then add
a stochastic stress tensor to the Stokes equation for the fluid velocity
\cite{BrownianBlobs,FBIM}. After elimination of the fluid one obtains
(\ref{eq:noiseSystem}). This can most simply be done by starting
from the fully discrete saddle-point system (25) in \cite{RigidMultiblobs},
adding the stochastic stress tensor as done in \cite{BrownianBlobs},
and the using standard Schur complement techniques to eliminate the
fluid velocity and pressure.}, to get the saddle-point linear system
\begin{equation}
\begin{bmatrix}\Mob & -\K\\
-\K^{T} & \V 0
\end{bmatrix}\begin{bmatrix}\V{\lambda}\\
\V U
\end{bmatrix}=\begin{bmatrix}-\sqrt{2k_{B}T/\D t}\;\Mob^{1/2}\V W\\
\V 0
\end{bmatrix}.\label{eq:noiseSystem}
\end{equation}
The solution of this system can be written using equation (\ref{eq:sysDense})
as 
\begin{equation}
\V U=\sqrt{2k_{B}T/\D t}\;\N\K^{T}\Mob^{-1}\Mob^{1/2}\V W=\sqrt{2k_{B}T/\D t}\;\N\K^{T}\Mob^{-1/2}\V W.\label{eq:noiseDense}
\end{equation}
It is not hard to see that we can identify the matrix 
\begin{equation}
\N^{1/2}\equiv\N\K^{T}\Mob^{-1/2}\label{eq:sqrtN_def}
\end{equation}
as a ``square root'' of the mobility, since 
\begin{align}
\N^{1/2}\left(\N^{1/2}\right)^{T} & =\N\K^{T}\Mob^{-1/2}\left(\Mob^{-1/2}\right)^{T}\K\N\\
 & =\N\left(\K^{T}\Mob^{-1}\K\right)\N=\N\N^{-1}\N=\N.
\end{align}
Thus, equation (\ref{eq:noiseDense}) becomes 
\begin{equation}
\V U=\sqrt{\frac{2k_{B}T}{\D t}}\;\N^{1/2}\V W.
\end{equation}
Hence the Brownian ``velocities'' (more precisely, the Brownian
displacements $\V U\D t$) for the rigid bodies can be computed by
solving a mobility problem with random slip given by Brownian velocities
for the blobs. Observe that we need only a \emph{single} application
of $\Mob^{1/2}$ to a vector, and the solution of a \emph{single}
mobility problem, to compute \emph{both} the deterministic and the
Brownian increments (but not including the stochastic drift terms
yet). Note that the same construction of $\N^{1/2}$ is used in the
recently-developed FBIM \cite{FBIM}, with the slip velocity $\slip$
interpreted as a random surface velocity distribution with covariance
equal to the Green's function for periodic Stokes flow (i.e., the
periodic Stokeslet).

Given an efficient routine to compute the product $\Mob\V{\lambda}$
for a given $\V{\lambda}$, as discussed in section \ref{sec:multN},
a preconditioned Lanczos-type iterative method to compute the product
$\Mob^{1/2}\V W$ was proposed in \cite{SquareRootPreconditioning}.
In unbounded or periodic domains the number of iterations increases
with the size of $\Mob$, and the cost of computing $\Mob^{1/2}\V W$
is many times than that of computing $\Mob\V{\lambda}$. In the PSE
method an additional splitting of $\Mob$ into a near-field and far-field
components is introduced, and the Lanczos method is only applied to
the near field, while the far-field component is handled using fluctuating
hydrodynamics. For particles confined close to a no-slip wall, the
friction with the floor screens the hydrodynamic interactions to decay
like inverse distance cubed. This makes the Lanczos iteration converge
in a small number of iterations independent of the number of blobs
\cite{MagneticRollers}. However, for rigid multiblobs the number
of iterations is higher than for single blobs because of the increased
ill-conditioning of $\Mob$ due to the presence of (nearly-)touching
blobs.

In this work, we employ a block diagonal preconditioner $\aMob\approx\Mob$
for the Lanczos algorithm \cite{SquareRootPreconditioning} that substantially
reduces the number of iterations in the computation of $\Mob^{1/2}\V W$
for rigid multiblobs. Similar block-diagonal or diagonal preconditioners
have been used for Stokesian suspensions by other authors \cite{SPME_Fibers,SD_SpectralEwald,BoundaryIntegral_SpheroidQBX,ForceCoupling_Fluctuations,RigidMultiblobs_Swan}.
In the preconditioner, which is also used to solve the saddle-point
system (\ref{eq:saddle}) \cite{RigidMultiblobs}, we ignore hydrodynamic
interactions between \emph{distinct} bodies $p$ and $q$, $\widetilde{\M{\mathcal{M}}}^{(pq)}=\delta_{pq}\M{\mathcal{M}}^{(pp)}$.
For each body $p$ we explicitly form a dense blob-blob mobility matrix
$\M{\mathcal{M}}^{(pp)}$ (equal to the diagonal block of $\M{\mathcal{M}}$
corresponding to body $p$), ignoring the presence of other bodies.
The preconditioner for the Lanczos method is a block diagonal matrix
$\M L=\aMob^{\frac{1}{2}}$ composed of the Cholesky factors of $\M{\mathcal{M}}^{(pp)}$.
We pre-compute $\M L$ once per time step (or less frequently if desired)
and then reuse it in the iterative solves in that time step.

\begin{figure}[h]
\centering{} \includegraphics[width=0.75\textwidth]{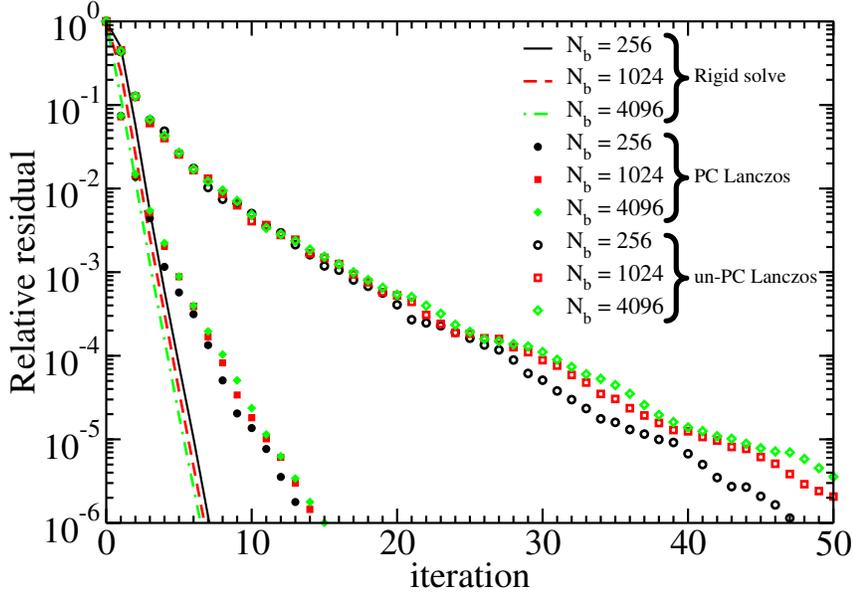}
\caption{\label{fig:BoomResidual}Convergence of iterative solvers for the
problem described in section \ref{sec:ManyBooms}, a suspension of
256, 1024 or 4096 colloidal boomerangs (each containing 15 blobs)
sedimented near a bottom wall. Convergence of the preconditioned GMRES
iteration to solve equation (\ref{eq:saddle}), labeled as `Rigid
solve', is demarcated by solid lines. Convergence of the preconditioned
Lanczos method to compute $\protect\Mob^{1/2}\protect\V W$, labeled
as 'PC Lanczos', is demarcated by filled symbols, while the corresponding
results without preconditioning, labeled as 'un-PC Lanczos', are demarcated
by un-filled symbols.}
\end{figure}

In Fig. \ref{fig:BoomResidual} we probe the convergence of our preconditioned
solvers in a suspension of boomerang colloidal particles sedimented
over a rigid wall for surface area fraction $\phi\approx0.25$ (see
details in Sec. \ref{sec:Results}). The figure shows the number of
iterations required to reach a desired tolerance for both the solution
of (\ref{eq:saddle}), as well as for computing the product $\Mob^{1/2}\V W$,
for three different problem sizes. Also shown is the effect of preconditioning
on the convergence of the matrix root computation. We can see that
up to a relative tolerance of around $10^{-3}$, both the saddle point
solve, (\ref{eq:saddle}) and $\Mob^{1/2}\V W$ require roughly the
same number of iterations to converge, with the latter taking more
iterations when smaller tolerances are required. When preconditioning
is used, both iterative methods are shown to have convergence rates
independent of problem size. Nevertheless, computing matrix roots
represents another major bottleneck in integrating (\ref{Langevin})
for rigid multiblobs confined above a no-slip floor, with cost similar
to that of solving the saddle point system (\ref{eq:saddle}). Observe
that both the computation of the deterministic and the fluctuating
velocities involves repeated applications of $\Mob$, which dominates
the cost. Therefore, we seek to integrate (\ref{Langevin}) to a desired
accuracy in as few total number of applications of $\Mob$ as possible.

\section{\label{sec:RFD}Temporal Integrators and the Thermal Drift}

Our goal is to numerically integrate the overdamped Langevin equation
(\ref{Langevin}) as efficiently as possible. In section \ref{sec:BD}
we discussed efficient means of computing $\N\V F+\sqrt{2k_{B}T/\D t}\;\N^{1/2}\V W$,
all that remains is to find a way to efficiently generate the thermal
drift term $k_{B}T\ \partial_{\V Q}\cdot\N$. Capturing this drift
term is a common challenge in all methods for Brownian dynamics, and
the methods developed here are general and apply to any approach based
on solving mobility problems.

A widely-used method to capture $k_{B}T\ \partial_{\V Q}\cdot\N$
is due to Fixman \cite{BD_Fixman}, and can be seen as a midpoint
method to capture the Stratonovich product in a mixed Stratonovich-Ito
(also known as Klimontovich or kinetic interpretation \cite{KineticStochasticIntegral_Ottinger})
re-formulation of (\ref{Langevin}) \cite{BrownianMultiBlobs}. The
generalization of Fixman's method to account for particle orientations
is given in Section III of \cite{BrownianMultiBlobs}. The problem
with the Fixman scheme in the context of many-body suspensions is
that it requires the computation of $\N^{-1/2}\V W$. This is related
to solving resistance problems and is infeasible for many body simulations.
In particular, there is no known method to compute $\Mob^{-1}$ which
scales linearly with the problem size. Hence, the Fixman's scheme
has to be ruled out for use in many body simulations. Here we will
only use Fixman's method as a reference method for small problems
involving at most on the order of a hundred blobs, where dense linear
algebra is practicable \cite{BrownianMultiBlobs}.

In \cite{MultiscaleIntegrators,BrownianMultiBlobs}, some of us proposed
a means of capturing the drift term in (\ref{Langevin}) using a modification
of Fixman's approach. This idea, termed random finite difference (RFD),
is as follows. Given two Gaussian random vectors, $\Delta\V P$ and
$\Delta\V Q$, such that $\av{\Delta\V P\Delta\V Q^{T}}=\left(k_{B}T\right)\M I$,
the following relation holds for a configuration dependent matrix
$\sM B\left(\V Q\right)$, 
\begin{align}
 & \lim_{\delta\rightarrow0}\frac{1}{\delta}\av{\left\{ \sM B\left(\V Q+\delta\Delta\V Q\right)-\sM B\left(\V Q\right)\right\} \Delta\V P}=\label{eq:RFDoneside}\\
 & \lim_{\delta\rightarrow0}\frac{1}{\delta}\av{\left\{ \sM B\left(\V Q+\frac{\delta}{2}\Delta\V Q\right)-\sM B\left(\V Q-\frac{\delta}{2}\Delta\V Q\right)\right\} \Delta\V P}=\label{eq:RFDcenter}\\
 & \left\{ \partial_{\V Q}\sM B\left(\V Q\right)\right\} \colon\av{\Delta\V P\Delta\V Q^{T}}=k_{B}T\,\partial_{\V Q}\cdot\sM B\left(\V Q\right),\label{eq:RFDB}
\end{align}
where $\av{}$ denotes an ensemble average. In practice, we will implement
random finite differences by simply taking $\delta$ to be a \textit{small}
number. Thus we recognize, by analogy with standard finite differences,
equations (\ref{eq:RFDoneside}) and (\ref{eq:RFDcenter}) as one-sided
and centered approximations to (\ref{eq:RFDB}) with truncation errors
of $O\left(\delta\right)$ and $O\left(\delta^{2}\right)$ respectively.
Note that Fixman's scheme can be viewed as an RFD where $\delta=\sqrt{\Delta t}$,
$\sM B=\N$, $\Delta\V Q=\sqrt{k_{B}T}\N^{1/2}\V W$, and $\Delta\V P=\sqrt{k_{B}T}\N^{-1/2}\V W$
\cite{MultiscaleIntegrators}; see Section III.D in \cite{StokesianDynamics_Brownian}
for the first use of a $\delta$ independent of $\D t$ in order to
``avoid particle 'overlaps' in the intermediate configuration.''
A simpler choice, used in \cite{BrownianMultiBlobs,MagneticRollers},
is to take $\sM B=\N$, and use $\Delta\V P=\Delta\V Q=\sqrt{k_{B}T}\,\V W$.
Other more efficient choices have been constructed in a number of
specific contexts \cite{BrownianBlobs,PSE_Stresslets,FluctuatingFCM_DC}.
In order to best pick $\delta$, we must balance the truncation error
with other sources of error introduced from the inexact multiplication
of $\sM B$. At best, multiplication by $\sM B$ is calculated to
machine precision and $\delta$ may be taken to be quite small. At
worst, multiplication of $\sM B$ is only computed approximately to
within some relative tolerance $\epsilon$, as would be the case when
we take $\sM B=\N$ and matrix vector multiplications are computed
using the iterative method described in section (\ref{sec:multN}).
In this case, using one-sided differencing can lead to large truncation
errors when loose solver tolerances are used, and we recommend that
only central differencing be used. 

In \cite{BrownianMultiBlobs}, an Euler-Maruyama (EM) RFD (EM-RFD)
scheme is presented to solve (\ref{Langevin}), using a one-sided
RFD on $\N$. A scalable variant of this using a central RFD is a
trivial extension summarized in appendix \ref{add:EMrfd}, where we
clarify how to do this using iterative solvers and also with care
for different units for length and orientation. This scheme requires
\emph{three} solutions of the saddle point system (\ref{eq:saddle})
and one application of $\Mob^{1/2}$ per timestep, and is only first-order
accurate even deterministically. By using different choices for $\Delta\V P$
and $\Delta\V Q$ in (\ref{eq:RFDcenter}), we will reduce the cost
of capturing the stochastic drift term considerably. In section \ref{Traction}
we will present an EM Traction (EM-T) scheme which only requires \emph{two}
solutions of the saddle point system. The trapezoidal slip (T-S) scheme
presented in section \ref{sec:Slip} still requires three solutions
of the saddle point system but achieves higher accuracy, notably,
it is second-order accurate deterministically just like the Fixman
midpoint scheme given in \cite{BrownianMultiBlobs}. We will empirically
compare these two schemes in terms of accuracy per computational effort
in Section \ref{sec:ManyBooms}.

\subsection{\label{Traction}Euler-Maruyama Traction (EM-T) Scheme}

To improve the efficiency of the scheme given in appendix \ref{add:EMrfd},
we propose a different means of computing the drift term. Using the
chain rule, we can split the divergence of the body mobility matrix
into three pieces, 
\begin{align}
 & \partial_{\V Q}\cdot\N=-\N\left(\partial_{\V Q}\N^{-1}\right)\colon\N=-\N\left(\partial_{\V Q}\left\{ \K^{T}\Mob^{-1}\K\right\} \right)\colon\N=\label{DN}\\
 & -\N\left\{ \partial_{\V Q}\K^{T}\right\} \colon\Mob^{-1}\K\N-\N\K^{T}\left\{ \partial_{\V Q}\Mob^{-1}\right\} \colon\K\N-\N\K^{T}\Mob^{-1}\left\{ \partial_{\V Q}\K\right\} \colon\N=\nonumber \\
 & -\N\left\{ \partial_{\V Q}\K^{T}\right\} \colon\Mob^{-1}\K\N+\N\K^{T}\Mob^{-1}\left\{ \partial_{\V Q}\Mob\right\} \colon\Mob^{-1}\K\N-\N\K^{T}\Mob^{-1}\left\{ \partial_{\V Q}\K\right\} \colon\N.\nonumber 
\end{align}
where colon denotes contraction; this calculation is done more precisely
using index notation in appendix \ref{Add:Quat}. Unlike $\N$, we
can efficiently compute the action of $\K^{T}$, $\Mob$, and $\K$,
without the need for a linear solver \footnote{In this paper, we evaluate the action of $\Mob$ directly using a
summation on the GPU, which gives accuracy comparable to roundoff.
But other more scalable implementations of the action of the RPY mobility
would use approximate methods like the fast multipole method (FMM)
\cite{RPY_FMM,OseenBlake_FMM} or the Spectral Ewald method \cite{SpectralRPY},
which would have an input relative error tolerance of their own, larger
than the roundoff error.}. Thus, we can use a random finite difference to compute the three
derivatives $\partial_{\V Q}\K^{T}$, $\partial_{\V Q}\Mob$, and
$\partial_{\V Q}\K$ in equation (\ref{DN}) separately. When selecting
the value of $\delta$ for these computations, we must balance the
truncation error of the RFD (we will only consider centered differences)
with the relative accuracy in computing the product of the operator
(i.e $\K$, $\Mob$). If the matrix-vector products are computed directly,
then we balance the truncation error with the machine precision and
take $\delta\sim10^{-3}$ when single precision is used \footnote{On many GPUs it is more efficient to use single-precision arithmetic.},
and $\delta\sim10^{-6}$ for double precision. However, if we only
compute the action of $\Mob$ to within some relative accuracy $\epsilon$,
as would be the case if we used the FMM or PSE method, we must take
$\delta\sim\epsilon^{1/3}$.

To utilize equation (\ref{DN}), we first generate random forces and
torques for each body $p$ 
\begin{equation}
\V W_{p}^{FT}=k_{B}T\begin{bmatrix}L_{p}^{-1}\V W_{p}^{f}\\
\V W_{p}^{\tau}
\end{bmatrix},\label{eq:W_FT}
\end{equation}
where $\V W_{p}^{f},\,\V W_{p}^{\tau}$ are standard Guassian random
vectors, and $L_{p}$ is a measure of the body length. Note the choice
of length scale used in the blocks of (\ref{eq:W_FT}) is to minimize
the variance of the RFD estimate, as we explain in appendix \ref{add:EMrfd}.
We then solve a mobility problem with random applied forces and torques
$\V W^{FT}=\left[\V W_{p}^{FT}\right]$, for both the random traction
force $\V{\lambda}^{\text{RFD}}$, and the random rigid velocity $\V U^{\text{RFD}}$,
\begin{align}
\V{\lambda}^{\text{RFD}} & =\Mob^{-1}\K\N\ \V W^{FT}\\
\V U^{\text{RFD}} & =\N\ \V W^{FT}.
\end{align}
To compute the relevant random finite difference terms, we randomly
displace the particles to $\V Q^{\pm}$, where
\begin{equation}
\V Q_{p}^{\pm}=\V Q_{p}\pm\frac{\delta}{2}\Delta\V Q_{p}=\V Q_{p}\pm\frac{\delta}{2}\begin{bmatrix}L_{p}\V W_{p}^{f}\\
\V W_{p}^{\tau}
\end{bmatrix}.
\end{equation}
Using this and equation (\ref{DN}), we are able to compute the necessary
drift term using random finite differences as 
\begin{align}
\text{Drift}= & -\frac{1}{\delta}\N\left\{ \K^{T}(\V Q^{+})-\K^{T}(\V Q^{-})\right\} \ \V{\lambda}^{\text{RFD}}\label{eq:traction}\\
 & +\frac{1}{\delta}\N\K^{T}\Mob^{-1}\left\{ \Mob(\V Q^{+})-\Mob(\V Q^{-})\right\} \ \V{\lambda}^{\text{RFD}}\nonumber \\
 & -\frac{1}{\delta}\N\K^{T}\Mob^{-1}\left\{ \K(\V Q^{+})-\K(\V Q^{-})\right\} \ \V U^{\text{RFD}}\nonumber \\
\approx & \biggl(-\N\left\{ \partial_{\V Q}\K^{T}\right\} \Mob^{-1}\K\N\nonumber \\
 & +\N\K^{T}\Mob^{-1}\left\{ \partial_{\V Q}\Mob\right\} \Mob^{-1}\K\N\nonumber \\
 & -\N\K^{T}\Mob^{-1}\left\{ \partial_{\V Q}\K\right\} \N\biggl):\left[\V W^{FT}\ \left(\Delta\V Q\right)^{T}\right]\nonumber \\
= & \partial_{\V Q}\N:\left[\V W^{FT}\ \left(\Delta\V Q\right)^{T}\right],\nonumber 
\end{align}
where all operators and derivatives are evaluated at the same point
$\V Q$ unless otherwise noted and $\left(\Delta\V Q\right)^{T}$
denotes the transpose of $\Delta\V Q$. Hence, in expectation, we
have 
\begin{equation}
\left\langle \text{Drift}\right\rangle \approx k_{B}T\ \partial_{\V Q}\cdot\N.
\end{equation}
This computation is detailed in index notation, accounting for the
constrained quaternion representation of orientations, in Appendix
\ref{Add:tractionDense}.

To leading order in $\delta$, the method of computing the drift proposed
in equation (\ref{eq:traction}), termed the \textit{traction-corrected}
RFD, is equivalent to the direct RFD on $\N$ used in appendix \ref{add:EMrfd}
when exact linear algebra is used. However, using the traction-corrected
RFD allows the use of inexact, iterative mobility solvers, without
incurring additional restrictions on the small parameter $\delta$
from the prescribed solver tolerance. Furthermore, we are able to
capture the drift term in equation (\ref{Langevin}) with only two
saddle point solves rather than the three required if we were to use
an RFD on $\N$ directly. Our Euler-Maruyama Traction (EM-T) Scheme
is summarized in algorithm \ref{alg:traction}, and is analyzed in
Appendix \ref{Add:tractionDense}.

\begin{algorithm}[h]
\caption{\label{alg:traction} Euler-Maruyama Traction (EM-T) Scheme}
\begin{enumerate}
\item \label{step:TracStep1}Compute relevant quantities for capturing drift:

\begin{enumerate}
\item Form $\V W^{FT}=\left[\V W_{p}^{FT}\right],$ where 
\[
\V W_{p}^{FT}=k_{B}T\begin{bmatrix}L_{p}^{-1}\V W_{p}^{f}\\
\V W_{p}^{\tau}
\end{bmatrix}
\]
and $\V W_{p}^{f},\,\V W_{p}^{\tau}$ are standard Gaussian random
vectors.
\item \label{step:ULam}Solve RFD mobility problem:
\[
\begin{bmatrix}\Mob^{n} & -\K^{n}\\
-\left(\K^{T}\right)^{n} & \V 0
\end{bmatrix}\begin{bmatrix}\V{\lambda}^{\text{RFD}}\\
\V U^{\text{RFD}}
\end{bmatrix}=\begin{bmatrix}0\\
-\V W^{FT}
\end{bmatrix}.
\]
\item \label{step:URFD}Randomly displace particles to:
\[
\V Q_{p}^{\pm}=\V Q_{p}^{n}+\frac{\delta}{2}\begin{bmatrix}L_{p}\V W_{p}^{f}\\
\V W_{p}^{\tau}
\end{bmatrix}.
\]
\item \label{step:DFDS}Compute the force-drift, $\V D^{F}$, and the slip-drift,
$\V D^{S}$:
\begin{align*}
\V D^{F} & =\frac{1}{\delta}\left\{ \K^{T}\left(\V Q^{+}\right)-\K^{T}\left(\V Q^{-}\right)\right\} \V{\lambda}^{\text{RFD}}\\
\V D^{S} & =\frac{1}{\delta}\left\{ \Mob\left(\V Q^{+}\right)-\Mob\left(\V Q^{-}\right)\right\} \V{\lambda}^{\text{RFD}}-\frac{1}{\delta}\left\{ \K\left(\V Q^{+}\right)-\K\left(\V Q^{-}\right)\right\} \V U^{\text{RFD}}.
\end{align*}
Note that different $\delta$ may be used for the RFDs on $\K$ and
$\Mob$ depending on the relative accuracy with which the action of
$\Mob$ is evaluated.
\end{enumerate}
\item \label{step:MhalfEM}Compute $\left(\Mob^{1/2}\right)^{n}\V W^{n}$
using a preconditioned Lancoz method or PSE.
\item \label{step:Udef}Evaluate forces and torques at $\V F^{n}=\V F\left(\V Q^{n},\,t\right)$
and solve the mobility problem:
\[
\begin{bmatrix}\Mob^{n} & -\K^{n}\\
-\left(\K^{T}\right)^{n} & \V 0
\end{bmatrix}\begin{bmatrix}\V{\lambda}^{n}\\
\V U^{n}
\end{bmatrix}=\begin{bmatrix}-\V D^{S}-\sqrt{2k_{B}T/\D t}\left(\Mob^{1/2}\right)^{n}\V W^{n}\\
-\V F^{n}+\V D^{F}
\end{bmatrix}.
\]
\item \label{step:U} Update configurations to time $t+\D t$:
\[
\V Q^{n+1}=\V Q^{n}+\Delta t\V U^{n}.
\]
\end{enumerate}
\end{algorithm}

\subsection{\label{sec:Slip}Trapezoidal Slip (T-S) Scheme}

In section \ref{Traction}, we developed a method to efficiently and
accurately generate the necessary drift term in an Euler-Maruyama
scheme. However, when second order deterministic accuracy is desired,
we may wish to use a midpoint or trapezoidal scheme \cite{MultiscaleIntegrators}.
Some higher order methods, however, will generate additional drift
terms due to the Brownian increment being evaluated at multiple time
levels. As an example, consider a naive two-solve implementation of
the trapezoidal scheme:
\begin{align}
\widetilde{\V Q}= & \V Q^{n}+\D t\N^{n}\V F^{n}+\sqrt{2\D tk_{B}T}\left(\N\K^{T}\Mob^{-1}\right)^{n}\left(\Mob^{1/2}\right){}^{n}\ \V W^{n}\label{eq:IncTrap1}\\
\V Q^{n+1}= & \V Q^{n}+\frac{\D t}{2}\left(\N^{n}\V F^{n}+\aN\widetilde{\V F}\right)\label{eq:IncTrap2}\\
+ & \sqrt{\frac{\D tk_{B}T}{2}}\left\{ \left(\N\K^{T}\Mob^{-1}\right)^{n}+\aN\aK^{T}\aMob^{-1}\right\} \left(\Mob^{1/2}\right){}^{n}\ \V W^{n},\nonumber 
\end{align}
where superscripts and tildes indicate the point at which quantities
are evaluated, e.g., $\aN\equiv\N\left(\widetilde{\V Q}\right)$. 

As shown in Appendix \ref{Add:SlipDense}, the thermal drift produced
by the final velocity update in equation (\ref{eq:IncTrap2}) (in
expectation) is 
\begin{equation}
\left\langle \text{\text{Drift part 1}}\right\rangle =\av{\frac{\V Q^{n+1}-\V Q^{n}}{\D t}}\approx\left(k_{B}T\right)\ \N\K^{T}\Mob^{-1}\left\{ \partial_{\V Q}\K\right\} \colon\N,\label{eq:trapdrift}
\end{equation}
We recognize this as the third term in equation (\ref{DN}) and hence,
we may use it to generate the full, desired drift. Examining equation
(\ref{DN}) reveals that we must generate the remaining two terms
\begin{equation}
-\N\left\{ \partial_{\V Q}\K^{T}\right\} \colon\Mob^{-1}\K\N+\N\K^{T}\Mob^{-1}\left\{ \partial_{\V Q}\Mob\right\} \colon\Mob^{-1}\K\N,\label{eq:TrapTerms}
\end{equation}
in order to capture the desired drift.

In section \ref{Traction}, we generated random traction forces of
the form $\Mob^{-1}\K\N\V W^{FT}$, and used these as the $\D{\V P}$
in equation (\ref{eq:RFDB}) to compute a traction-corrected RFD approximation
to (\ref{eq:TrapTerms}). Here we propose a different \emph{slip-corrected}
RFD method to compute the two terms in (\ref{eq:TrapTerms}). For
each body, we generate a vector of random blob displacements $\slipW^{D}=\left[L_{p}\V W_{p}^{s}\right]$,
and random blob forces $\slipW^{F}=\left[\frac{k_{B}T}{L_{p}}\V W_{p}^{s}\right],$
where $L_{p}$ is a length scale for body $p$, and $\V W^{s}$ is
a random Gaussian vector. We may then compute rigid body displacements,
$\D{\V Q^{\text{RFD}}}=\N\K^{T}\Mob^{-1}\slipW^{D}$, which may be
used as $\D{\V Q}$ in equation (\ref{eq:RFDB}) to compute an RFD
approximation to (\ref{eq:TrapTerms}). That is, we may approximate
the missing drift terms (\ref{eq:TrapTerms}) by computing

\begin{align}
\text{Drift part 2}= & -\frac{1}{\delta}\N\left\{ \K^{T}(\V Q^{+})-\K^{T}(\V Q^{-})\right\} \ \slipW^{F}\label{eq:slip}\\
 & +\frac{1}{\delta}\N\K^{T}\Mob^{-1}\left\{ \Mob(\V Q^{+})-\Mob(\V Q^{-})\right\} \ \slipW^{F}\nonumber \\
\approx & \left(-\N\left\{ \partial_{\V Q}\K^{T}\right\} +\N\K^{T}\Mob^{-1}\left\{ \partial_{\V Q}\Mob\right\} \right)\colon\left[\slipW^{F}\ \left(\D{\V Q^{\text{RFD}}}\right)^{T}\right]\nonumber \\
= & \left(k_{B}T\right)\left(-\N\left\{ \partial_{\V Q}\K^{T}\right\} +\N\K^{T}\Mob^{-1}\left\{ \partial_{\V Q}\Mob\right\} \right)\colon\Mob^{-1}\K\N\left[\V W^{s}\left(\V W^{s}\right)^{T}\right],\nonumber 
\end{align}
where as before $\V Q_{p}^{\pm}=\V Q_{p}\pm\frac{\delta}{2}\D{\V Q_{p}^{\text{RFD}}}$.
Hence, in expectation we obtain the missing drift terms (\ref{eq:TrapTerms}),
\[
\left\langle \text{\text{Drift part 2}}\right\rangle \approx k_{B}T\left(-\N\left\{ \partial_{\V Q}\K^{T}\right\} \colon\Mob^{-1}\K\N+\N\K^{T}\Mob^{-1}\left\{ \partial_{\V Q}\Mob\right\} \colon\Mob^{-1}\K\N\right),
\]
which combined with (\ref{eq:trapdrift}) gives us the desired drift
$k_{B}T\ \partial_{\V Q}\cdot\N$.

Our Trapezoidal Slip (T-S) scheme is summarized in algorithm \ref{alg:sliptrap},
and is analyzed in Appendix \ref{Add:SlipDense}. It involves three
mobility solves and one Lanczos computation per time step, just like
the EM-RFD scheme given in Algorithm \ref{alg:RFDN}, however, T-S
is second order deterministically just like the Fixman midpoint scheme.

\begin{algorithm}[h]
\caption{\label{alg:sliptrap} Trapezoidal Slip (T-S) scheme}
\begin{enumerate}
\item Compute relevant quantities for capturing drift:

\begin{enumerate}
\item Generate random Gaussian directions $\V W^{s}$ for each blob, and
form the composite vectors of blob displacements $\slipW^{D}=\left[L_{p}\V W_{p}^{s}\right]$
and blob forces $\slipW^{F}=\left[\frac{k_{B}T}{L_{p}}\V W_{p}^{s}\right]$.
\item \label{step:UrfdT}Solve RFD mobility (more precisely, displacement)
problem:
\[
\begin{bmatrix}\Mob^{n} & -\K^{n}\\
-\left(\K^{T}\right)^{n} & \V 0
\end{bmatrix}\begin{bmatrix}\V{\lambda}^{\text{RFD}}\\
\D{\V Q^{\text{RFD}}}
\end{bmatrix}=\begin{bmatrix}-\slipW^{D}\\
0
\end{bmatrix}.
\]
\item \label{step:QpmT}Randomly displace particles to $\V Q^{\pm}$:
\[
\V Q^{\pm}=\V Q^{n}\pm\frac{\delta}{2}\D{\V Q^{\text{RFD}}}
\]
\item \label{step:DFDS2}Compute the force-drift, $\V D^{F}$, and the slip-drift,
$\V D^{S}$, where:
\begin{align*}
\V D^{F} & =\frac{1}{\delta}\left\{ \K^{T}\left(\V Q^{+}\right)-\K^{T}\left(\V Q^{-}\right)\right\} \slipW^{F},\\
\V D^{S} & =\frac{1}{\delta}\left\{ \Mob\left(\V Q^{+}\right)-\Mob\left(\V Q^{-}\right)\right\} \slipW^{F}.
\end{align*}
Note that different $\delta$ may be used for the two RFDs depending
on the relative accuracy with which the action of $\Mob$ is evaluated.
\end{enumerate}
\item Compute $\left(\Mob^{1/2}\right)^{n}\V W^{n}$ using a preconditioned
Lancoz method or PSE.
\item \label{step:unT}Evaluate forces and torques at $\V F^{n}=\V F\left(\V Q^{n},\,t\right)$
and solve predictor mobility problem:
\[
\begin{bmatrix}\Mob^{n} & -\K^{n}\\
-\left(\K^{T}\right)^{n} & \V 0
\end{bmatrix}\begin{bmatrix}\V{\lambda}^{n}\\
\V U^{n}
\end{bmatrix}=\begin{bmatrix}-\sqrt{2k_{B}T/\D t}\left(\Mob^{1/2}\right)^{n}\V W^{n}\\
-\V F^{n}
\end{bmatrix}.
\]
\item \label{step:Qtil}Update configurations to predicted position $\widetilde{\V Q}$:
\[
\widetilde{\V Q}=\V Q^{n}+\D t\V U^{n}.
\]
\item \label{step:unp1T}Evaluate forces and torques at $\widetilde{\V F}=\V F\left(\widetilde{\V Q},\,t\right)$
and solve corrector mobility problem at the predicted position $\widetilde{\V Q}$:
\[
\begin{bmatrix}\aMob & -\aK\\
-\aK^{T} & \V 0
\end{bmatrix}\begin{bmatrix}\widetilde{\V{\lambda}}\\
\widetilde{\V U}
\end{bmatrix}=\begin{bmatrix}-2\V D^{S}-\sqrt{\frac{2k_{B}T}{\Delta t}}(\Mob^{1/2})^{n}\V W^{n}\\
-\widetilde{\V F}+2\V D^{F}
\end{bmatrix}.
\]
\item \label{step:unun1}Update configurations to corrected position $\V Q^{n+1}$:
\[
\V Q^{n+1}=\V Q^{n}+\frac{\D t}{2}\left(\V U^{n}+\widetilde{\V U}\right).
\]
\end{enumerate}
\end{algorithm}

It is important to point out that by using either the traction-corrected
RFD (i.e., applying random uncorrelated forces and torques on the
particles) or the slip-corrected RFD (i.e., applying random uncorrelated
slip on the particles' surfaces), one can construct a multitude of
schemes that give the desired stochastic drift term in expectation
for sufficiently small $\D t$. For example, an alternative method
to generate the remaining drift terms in (\ref{eq:TrapTerms}), while
still using the trapezoidal rule, would be to compute $\V D^{S}$
and $\V D^{F}$ analogous to step \ref{step:TracStep1} of algorithm
\ref{alg:traction} but without the term involving $\V U^{\text{RFD}}$
(which is already included via the trapezoidal corrector step). In
numerical tests, we have found such a Trapezoidal Traction (T-T) scheme
to perform very similarly to the (T-S) scheme for all solver tolerances
and time step sizes examined. We also examined midpoint variants \cite{DFDB}
of the (T-T) and (T-S) schemes, both of which require an additional
application of $\Mob^{1/2}$ in the corrector step, and found them
to be inferior in terms of cost-accuracy balance than the T-S scheme
presented here \footnote{The midpoint slip scheme was much more robust than the midpoint traction
scheme for larger $\D t$.}. Modifications of the ideas presented in section \ref{Traction}
can be used to write a second order Adams-Bashforth traction (AB-T)
scheme, in which the second-order AB multistep rule is used for the
deterministic terms, while the drift term is computed analogously
to the EM-T scheme \cite{MagneticRollers}. We found, however, that
the AB-T scheme was inferior in accuracy compared to the T-S scheme,
especially for larger $\D t$.

\section{Results\label{sec:Results}}

In this section, we study the accuracy and efficiency of the numerical
schemes presented in sections \ref{Traction} and \ref{sec:Slip}
for suspensions of rigid particles sedimented above a no-slip bottom
wall. As mentioned in section \ref{sec:Nhalf}, the bottom wall acts
to screen the hydrodynamic interactions, thereby reducing the number
of iterations required for iterative methods to converge to a desired
tolerance \cite{MagneticRollers,RigidMultiblobs}. To prevent unphysical
particle overlaps with the wall due to the Brownian motion, we include
a soft repulsive wall-particle potential, and employ a regularized
form of the blob-blob mobility which ensures that $\Mob$ is SPD and
physical even when some blobs overlap the wall \cite{MagneticRollers}.

\begin{figure}[h]
\begin{centering}
\includegraphics[width=0.9\textwidth]{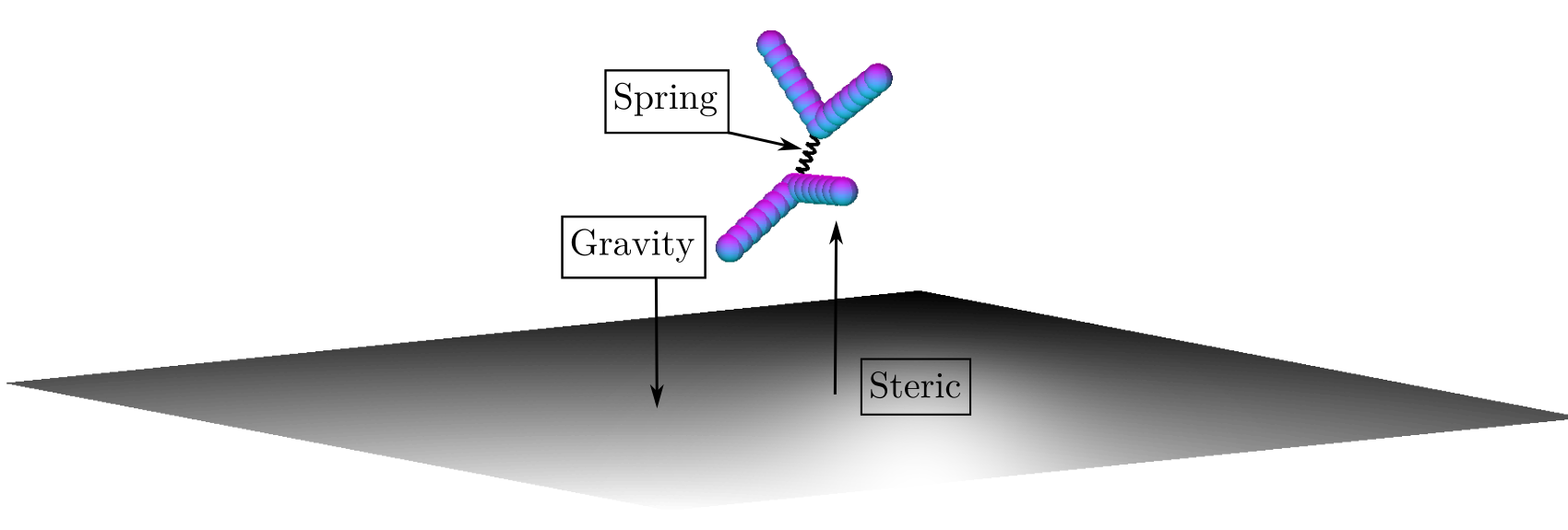}
\par\end{centering}
\begin{centering}
\includegraphics[width=1\textwidth]{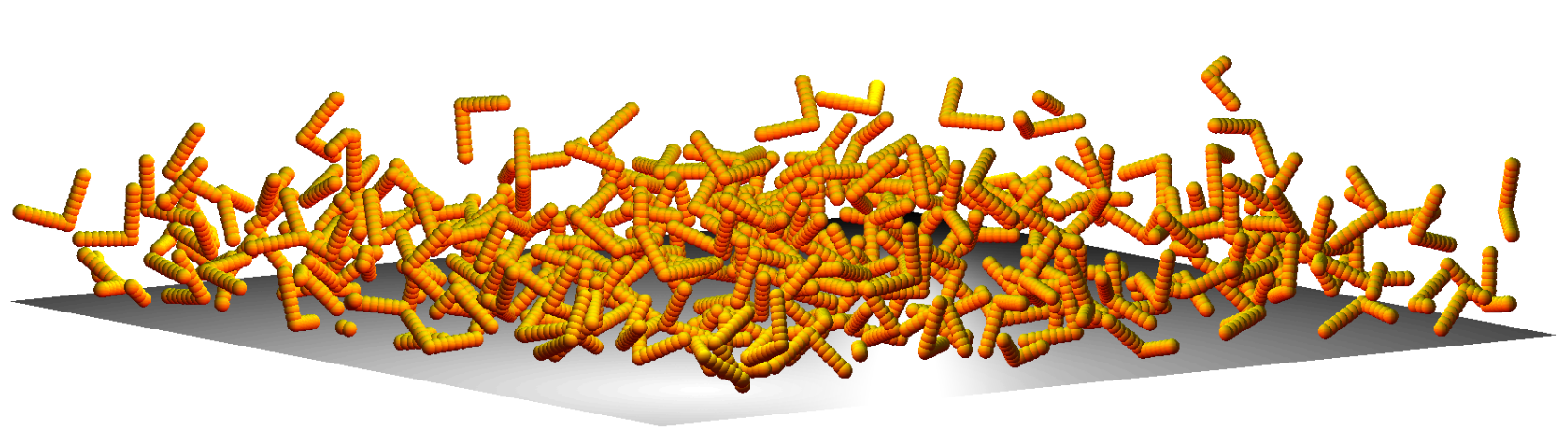}
\par\end{centering}
\centering{}\caption{\label{fig:Booms}Illustrations of the test problems involving colloidal
boomerangs. (Top panel) Sample configuration of a boomerang dimer
for the numerical experiments conducted in section \ref{sec:TwoBooms}.
(Bottom panel) Sample configuration of a boomerang suspension for
the numerical experiments conducted in section \ref{sec:ManyBooms}.
The shaded area is the part of the bottom wall that belongs to the
central unit cell used for the pseudo-periodic boundary conditions.}
\end{figure}

In sections \ref{subsec:WeakAccuracy} and \ref{sec:ManyBooms}, we
investigate the weak accuracy of our methods on suspensions of colloidal
right-angle \textquotedbl{}boomerangs\textquotedbl{}. Colloidal boomerangs
have been manufactured using lithography \cite{BoomerangDiffusion},
and the diffusion of a single boomerang above a wall was studied numerically
in \cite{BrownianMultiBlobs}. We will model a colloidal boomerang
as an L-shaped body composed of 15 blobs, with each `arm' of the boomerang
being composed of 7 blobs in straight line, plus a common eighth blob
shared by both arms, see Fig. \ref{fig:Booms}. In \cite{BrownianMultiBlobs}
it was found that blobs centers should be spaced approximately a distance
of $a$ apart, where $a$ is the hydrodynamic radius of a blob. Although
geometrically simple, boomerangs do not have spherical, axial, or
skew symmetry and therefore proper treatment of orientations is essential
to correctly model colloidal diffusion \cite{BrownianMultiBlobs}.
In subsection \ref{sec:TwoBooms}, we study two colloidal boomerangs
connected by an elastic string. The small problem size allows us to
reduce sampling (statistical) errors enough to accurately measure
temporal accuracy, and also to compare the schemes developed in this
work to Fixman's scheme \cite{BD_Fixman}, which requires dense linear
algebra to be used. In subsection \ref{sec:ManyBooms} we examine
a suspension of many boomerangs to more effectively assess the accuracy-efficiency
tradeoff for the schemes developed here.

In section \ref{sec:BrownRoll}, we revisit some of the computational
investigations reported in \cite{MagneticRollers,NonlocalShocks_Rollers}
for active suspensions of rotating colloids \cite{Rollers_NaturePhys,NonlocalShocks_Rollers}.
In these suspensions thermal motion sets the equilibrium gravitational
height of the colloids, and it is necessary to include Brownian motion
to enable quantitative comparisons to experiments \cite{MagneticRollers,NonlocalShocks_Rollers}.
At the same time, previous studies \cite{MagneticRollers,NonlocalShocks_Rollers}
used a minimally-resolved representation of the hydrodynamics, with
each particle represented by a single blob. This is not quantitatively
accurate when the microrollers are close to the wall or other colloids,
as in recent experiments \cite{Rollers_NaturePhys,NonlocalShocks_Rollers}.
We represent the spheres using either 12 or 42 blobs \cite{RigidMultiblobs}
in order to improve the accuracy of the hydrodynamic interactions,
and choose $a$ as roughly half the distance between vertices in the
multiblob sphere model following the recommendations in Sections IV
and V of \cite{RigidMultiblobs}.

\subsection{\label{subsec:WeakAccuracy}Weak accuracy for a dimer of boomerangs}

In this section we examine the equilibrium dynamics of a boomerang
dimer formed from two colloidal boomerangs connected by a harmonic
spring, as illustrated in the top panel of Fig. \ref{fig:Booms}.
The form of the potential connecting the two boomerangs is 
\begin{equation}
U_{\text{\text{spring}}}(d)=\frac{\kappa}{2}\left(d-l_{0}\right)^{2},
\end{equation}
where $d$ is the distance between the two boomerang's ``cross-points''
(the elbow of the L-shaped bodies), $l_{0}$ is a rest length, and
$\kappa$ is the spring constant. The steric interactions of the individual
blobs are captured through a repulsive Yukawa potential, 
\begin{equation}
U_{\text{steric}}(r)=\gamma\frac{e^{-r/b}}{r},
\end{equation}
where $b$ is the Debye length and $\gamma$ is the repulsion strength.
This Yukawa potential potential is also used for the steric interactions
of the bodies with the wall. In this section, we take $l_{0}=1\mu\text{m}$,
$\gamma=\kappa=0.096\frac{\text{m\text{g}}}{\text{s}^{2}}$, and $b=0.162\mu\text{m}$.
We will take the blob radius to be $a=0.324\mu$m and each blob will
have a buoyant (excess) mass $m_{e}=1.57\times10^{-11}$mg, giving
a net gravitational force $m_{e}g$ on each blob, where $g=9.81\frac{\text{m}}{\text{s}^{2}}$.
The total force and torque on each body are computed by adding contributions
of the spring, gravity, and steric repulsion over all the blobs comprising
the body \cite{RigidMultiblobs}. The bodies are suspended in water,
$\eta=1$ mPa$\cdot$s, at approximately room temperature, $T=300$K.
In these investigations, we nondimensionalize the time step using
the diffusive time scale for a single blob,
\[
\D{\tau}=\frac{k_{B}T}{6\pi\eta a^{3}}\D t.
\]

In the absence of non-conservative forces (i.e., for passive suspensions),
the equilibrium distribution for the particles' configuration is the
familiar Gibbs-Boltzmann (GB) distribution
\[
P_{\text{eq}}\left(\V Q\right)=P_{\text{GB}}\left(\V Q\right)=Z^{-1}\,\exp\left(-U\left(\V Q\right)/k_{B}T\right),
\]
where $U\left(\V Q\right)$ is the conservative potential from which
the external forces and torques are obtained. As demonstrated in our
prior work \cite{BrownianBlobs,BrownianMultiBlobs}, failure to consistently
include the stochastic drift term in BD simulations leads to strong
deviations from $P_{\text{GB}}\left(\V Q\right)$ in the presence
of confinement. Therefore, a strong test that our methods are consistent
with the overdamped Langevin equation (\ref{Langevin}) is to examine
how closely they reproduce (marginals of) the GB equilibrium distribution,
as we do in subsection \ref{sec:TwoBooms}. We use a Markov-chain
Monte Carlo (MCMC) method to very accurately sample the GB equilibrium
distribution and use this data to compute the error produced by each
scheme. At the same time, it is important to also confirm that our
schemes, unlike MCMC, correctly reproduce the \emph{dynamics} of the
particles even for time steps that are on the order of the diffusive
time scale, as we do in subsection \ref{sec:MSD}.

\subsubsection{\label{sec:TwoBooms}Static Accuracy}

\begin{figure}[h]
\centering{}\includegraphics[width=1\textwidth]{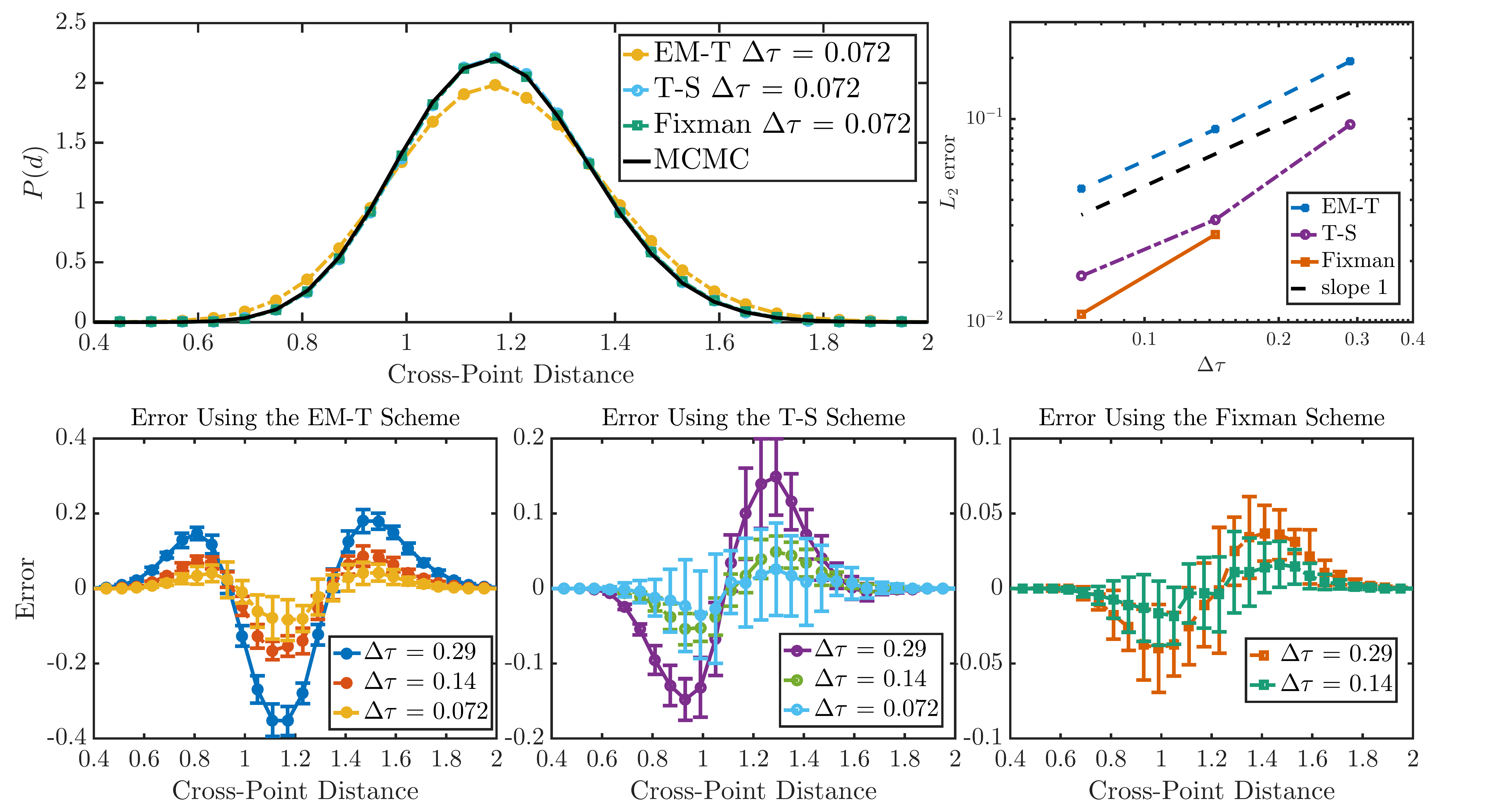}
\caption{\label{fig:CP}Numerical errors in the equilibrium distribution for
a boomerang dimer. (Upper left panel) Comparison between the correct
(marginal of the) Gibbs-Boltzmann distribution of the cross-point
distance, computed using an MCMC method, and numerical results from
the T-S and EM-T schemes for normalized time step size $\protect\D{\tau}=0.072$.
(Upper right panel) Cumulative error in the distribution of the cross-point
distance, as measured by the $L_{2}$ norm of the error in the histogram
$P(d)$, for the EM-T, T-S and Fixman schemes. (Lower panels) Error
in the distribution of the cross-point distance for the EM-T (left
panel), T-S (middle panel), and the Fixman scheme (right panel) for
several different time step sizes (see legend). Note that the scale
of the plots changes and that the Fixman scheme is unstable for $\protect\D{\tau}>0.29$.
Error bars indicate $95\%$ confidence intervals and are estimated
from multiple independent runs.}
\end{figure}

The stability limit for the EM-T scheme for the chosen parameters
was empirically estimated to be $\D{\tau}\lesssim0.3$. In Figure
\ref{fig:CP} we study how well our numerical methods reproduce selected
the Gibbs-Boltzmann equilibrium distribution for $\D{\tau=0.072,\,0.144,\,0.288}$.
We have examined a number of marginals of the equilibrium distribution,
but we focus here on the equilibrium distributions of the boomerang
cross-point to cross-point distance. We use a relative tolerance of
$10^{-4}$ in all iterative methods for the computations done in this
section. An investigation into the effect of solver tolerance on the
accuracy of the EM-T and T-S schemes showed no change in temporal
accuracy for all solver tolerances less than or equal to $10^{-3}$,
and overall accuracy was only slightly affected for solver tolerances
$10^{-3}-10^{-2}$, but then degraded rapidly for looser tolerances.
Note however, that using the same solver tolerance for all iterative
methods is perhaps not necessary to maintain temporal accuracy, and
looser tolerances may be used for the RFD-related linear solves. We
take the random finite difference parameter $\delta=10^{-6}$ for
both schemes as double precision was used for these calculations.
Results were obtained by averaging 20 independent realizations containing
$10^{5}$ samples, initialized from unique configurations sampled
from the equilibrium distribution using the MCMC algorithm. In addition
to the proposed EM-T and T-S schemes, we also compare with Fixman's
scheme given in Section III.B of \cite{BrownianBlobs}, implemented
using dense linear algebra.

Figure \ref{fig:CP} shows that even for the smallest time step size
considered, the T-S scheme is substantially more accurate than the
EM-T scheme. Note that no data is included for Fixman's scheme for
the largest time step size considered, because the scheme was seen
to be numerically unstable. While all of the schemes are seen to be
$\mathcal{O}(\Delta t)$ in the cummulative $L_{2}$ error, the order
constant for the T-S and Fixman schemes are clearly much lower than
the EM scheme. In terms of accuracy alone, the T-S scheme compares
very favorably with Fixman's scheme, while also enabling scalable
computations for suspensions of many rigid bodies.

\subsubsection{\label{sec:MSD}Dynamic Accuracy}

We now turn our attention to time-dependent statistics by examining
the equilibrium translational mean squared displacement (MSD) of the
cross point of one of the two connected boomerangs,
\[
\M D(t)=\avv{\D{\V q}_{p}(t)\left(\D{\V q_{p}}(t)\right)^{T}}=\avv{\left(\V q_{p}(t)-\V q_{p}(0)\right)\left(\V q_{p}(t)-\V q_{p}(0)\right)^{T}},
\]
where the average is an ensemble average over equilibrium trajectories
and $p=1$ or $p=2$. Since the trajectories of the two boomerangs
are statistically identical, we will average results over the two
particles to improve statistical accuracy. Here and in what follows
we will assume that the cross point is chosen as the tracking point
around which the boomerang rotates. We may define \emph{short-time}
and \emph{long-time} translational diffusion tensors,
\begin{equation}
\M{\chi}_{st}=\frac{1}{2}\,\lim_{t\to0}\frac{\M D(t)}{t},\hspace{1em}\M{\chi}_{lt}=\frac{1}{2}\,\lim_{t\to\infty}\frac{\M D(t)}{t}\label{eq:chi_st}
\end{equation}
respectively. The Stokes-Einstein relationship implies
\begin{equation}
\M{\chi}_{st}=\left(k_{B}T\right)\av{\N_{pp}^{(tt)}}_{GB},\label{eq:SE_short}
\end{equation}
where we have taken an average over the Gibbs-Boltzmann distribution
of the $3\times3$ translation-translation diagonal block of the mobility
matrix corresponding to body $p$. However, $\M{\chi}_{lt}$ admits
no such simple characterizations and is typically challenging to compute
accurately, requiring many samples from long simulations, as discussed
extensively in \cite{BrownianMultiBlobs}.

Since we are investigating diffusion near an infinite wall (placed
at $z=0$ with normal in the positive $z$ direction), under the influence
of gravity, we may define the parallel ($D_{p}^{\Vert}$) and perpendicular
($D_{p}^{\bot}$) MSD of body $p$ as

\begin{align}
D_{p}^{\Vert}(t)= & \M D_{xx}(t)+\M D_{yy}(t),\qquad D_{p}^{\bot}(t)=\M D_{zz}(t).\label{eq:parPerp}
\end{align}
At long times the perpendicular MSD $D_{p}^{\bot}(t)$ asymptotically
tends towards a finite value, related to the gravitational height
of the body \cite{BrownianMultiBlobs}. We focus here on the parallel
MSD $D_{p}^{\Vert}(t)$ as this is typically what is measured in experiments
\cite{BoomerangDiffusion,AsymmetricBoomerangs}.

At short times, we can use the Stokes-Einstein formula $D_{p}^{\Vert}(t)=2\left(k_{B}T\right)\av{\N_{11}^{(xx)}}_{GB}\,t$
to validate our simulations. To estimate the long-time MSD, we use
a non-equilibrium method based on linear response theory \cite{Frenkel_Smit_book}.
Specifically, if we pull one of the boomerangs with a force $\V F=F\,\hat{\V x}$
applied to the cross (tracking) point, 
\begin{equation}
\left\langle x_{p}(t)-x_{p}(0)\right\rangle _{\V F}=-\frac{F}{k_{B}T}\int_{0}^{t}\left\langle x_{p}(0)\dot{x}_{p}(t-t')\right\rangle _{0}\mathrm{d}t'=\frac{F}{2k_{B}T}\left\langle (x_{p}(t)-x_{p}(0))^{2}\right\rangle _{0}.\label{eq:noneq-MSD}
\end{equation}
Here the average on the left hand side is an average over nonequilibrum
trajectories initialized from the GB distribution, while the average
on the right hand side is an average over equilibrium trajectories.
The formula (\ref{eq:noneq-MSD}) relates the MSD at equilibrium with
the mean displacement under a external force. The nonequilibrium method
offers better statistical accuracy at long times over computing the
MSD if the applied force $F$ is sufficiently large but still small
enough to remain in the linear-response regime (for the simulations
reported below the Péclet number is $\text{Pe}=L\,F/\left(k_{B}T\right)\approx0.5$,
where $L=2.1\mu\text{m}$ is the boomerang arm length). To see this,
consider a one dimensional diffusion process with constant mobility
$\mu$,
\[
\frac{\mathrm{d}x(t)}{\mathrm{d}t}=\mu F+\sqrt{2k_{B}T\mu}\;\mathcal{W}(t),
\]
whose solution has mean $\av{x(t)}=\mu Ft$ and standard deviation
$\sqrt{2k_{B}T\mu t}$. The relative statistical uncertainty in the
mean displacement $\av{x(t)}$ is $\sqrt{2k_{B}T/(\mu F^{2}t)}$,
and therefore decays as time grows. By contrast, in the absence of
the force the mean MSD is $2k_{B}T\mu t$ while the standard deviation
of the MSD is $\sqrt{3}\left(2k_{B}T\mu t\right)$, and therefore
the relative statistical uncertainty in the MSD is independent of
time. This assumes we have an infinitely-long trajectory. In practice,
however, the finite length of the trajectories makes the MSD most
statistically accurate at short times, and it is beneficial to use
the nonequilibrium method to estimate the long-time diffusion coefficient.

\begin{figure}[h]
\centering{} \includegraphics[width=0.75\textwidth]{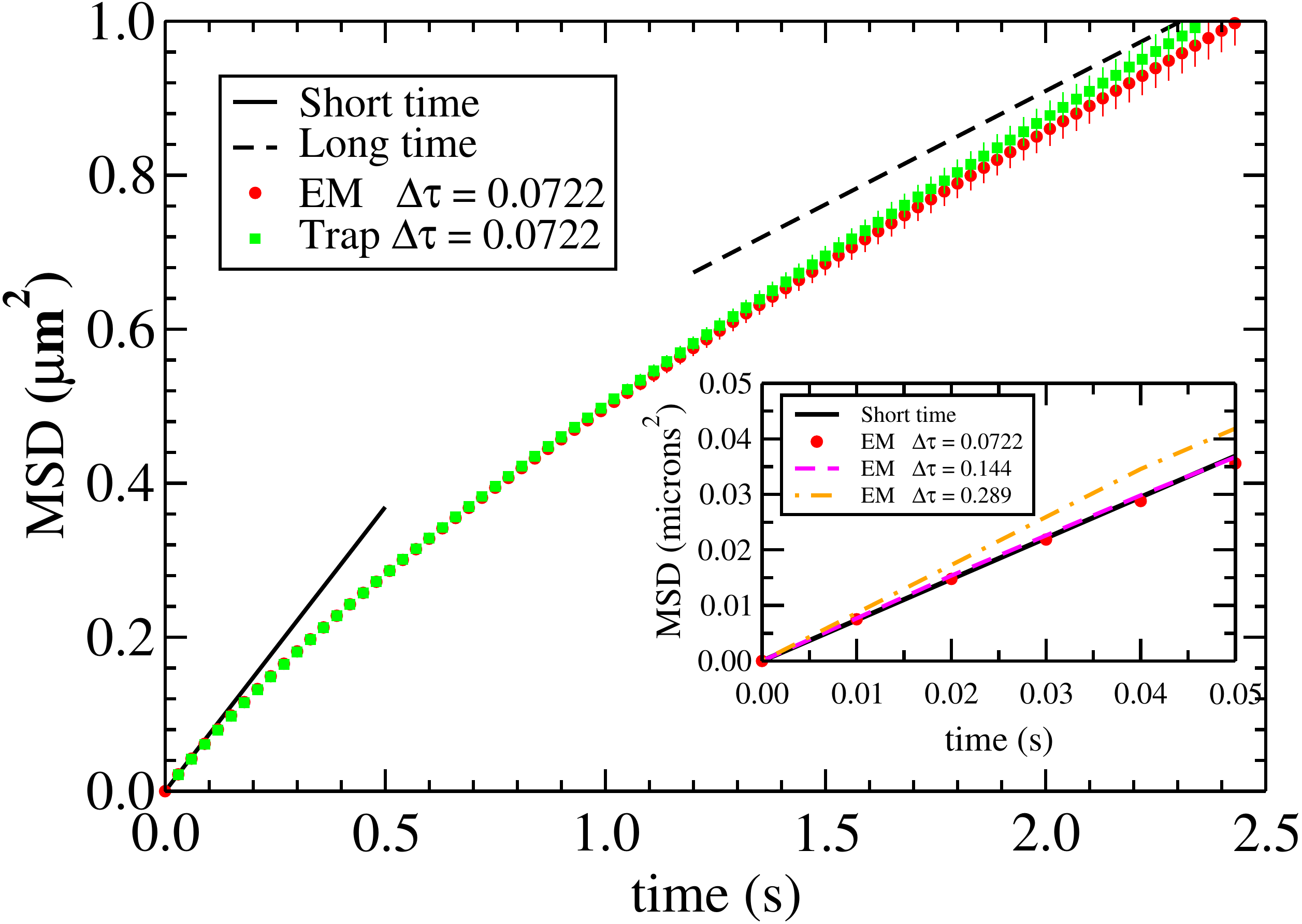}
\caption{\label{fig:TwoBoomsMSD}Mean square planar displacement of the cross
point of one of the boomerangs as obtained from trajectories sampled
using the EM-T and T-S schemes. Error bars indicate $95\%$ confidence
intervals. Black lines show the reference slope of the MSD at short
times (solid), as computed using the Stokes-Einstein formula, and
long times (dashed), as computed using a non-equilibrium method. (Inset)
Short time parallel MSD computed by the EM-T method for different
time step sizes.}
\end{figure}

Figure \ref{fig:TwoBoomsMSD} shows the results for the MSD of the
cross point of one of the boomerangs obtained using the EM-T and T-S
schemes. We can see that both schemes produce the correct slope of
the MSD at short times (short-time diffusion coefficient), as compared
with the Stokes-Einstein estimate obtained by computing $\av{\N_{11}^{(xx)}}_{GB}$
accurately using a Monte-Carlo method (solid black line). At long
times, to within statistical uncertainty, both schemes produce the
same slope of the MSD (long-time diffusion coefficient) as the non-equilibrium
method (dashed black line). Because the short time MSD is computed
as an equilibrium average over the Gibbs-Boltzmann distribution (see
(\ref{eq:SE_short})), the temporal accuracy with which a given scheme
samples the equilibrium GB distribution (as measured in section \ref{sec:TwoBooms})
directly effects the accuracy of the short time MSD. In particular,
the stochastic displacement produced by the EM-T scheme has covariance
proportional to $\D t\N(\V Q)$. Hence, the short time diffusion coefficient
produced by the EM-T scheme is independent of $\D t$, and the only
source of error stems from the error in the equilibrium distribution.
The inset of Fig. \ref{fig:TwoBoomsMSD} shows the short time MSD
produced by the EM-T scheme for different time steps. Here we see
clear improvement as the time step is reduced, analogous to the results
shown in Fig. \ref{fig:CP}. 

\subsection{\label{sec:ManyBooms}Accuracy and Efficiency for Many-Body Suspensions}

To compare the accuracy-efficiency tradeoff for the T-S scheme versus
the EM-T scheme, we investigate a dense suspension of freely-diffusing
colloidal boomerangs. All of the physical parameters and interaction
potentials of the simulation are taken to be the same as for the boomerang
dimer studied in Section (\ref{subsec:WeakAccuracy}), except that
we take the solver tolerance for all iterative solvers to be $10^{-3}$
as this was found to give statistically indistinguishable results
from tolerance $10^{-4}$. Further, to reduce the computation time,
we perform the arithmetic on the GPU (multiplication by $\Mob$) in
single precision and thus take random the finite difference parameter
$\delta=10^{-3}$ for both schemes. For suspensions we employ pseudo
periodic boundary conditions \cite{MagneticRollers}. Specifically,
for each blob $i$, we sum the hydrodynamic interactions with every
other blob $j$ (using the minimum image convention) and also the
8 nearest periodic images of blob $j$. This is fairly effective in
capturing the hydrodynamics of an infinitely periodic suspension,
without requiring involved Ewald summation techniques \cite{RegularizedStokeslets_WallPeriodic},
since the presence of the wall screens the hydrodynamic interactions
to decay as inverse \emph{cubed} distance. In order to challenge the
temporal integrators we consider a dense suspension in which steric
exclusion plays a large role in the dynamics.

Specifically, we will simulate 256 boomerangs in a domain which is
semi-infinite in $z$ and periodic with unit cell of length $45.3\mu$m
in both the $x$ and $y$ directions, as illustrated in the bottom
panel of Fig. \ref{fig:Booms}. We have examined a number of relevant
statistics (marginals of the equilibrium distribution) and found the
radial distribution function $g(r)$ to be the most sensitive measure.
We compute $g(r)$ using the minimum Euclidean distance between two
boomerangs, approximated as the smallest distance between a pair of
blobs taken from distinct bodies. We normalize $g(r)$ as if the suspension
were two dimensional, as was done in \cite{MagneticRollers}, in order
to ensure that $g(r)\rightarrow1$ for large $r$. For each scheme
and value of $\D{\tau}$, we simulate 8 independent trajectories with
$10^{4}$ samples in each, initialized using unique configurations
sampled from the equilibrium distribution using an MCMC algorithm.

\begin{figure}[h]
\centering{}\includegraphics[width=1\textwidth]{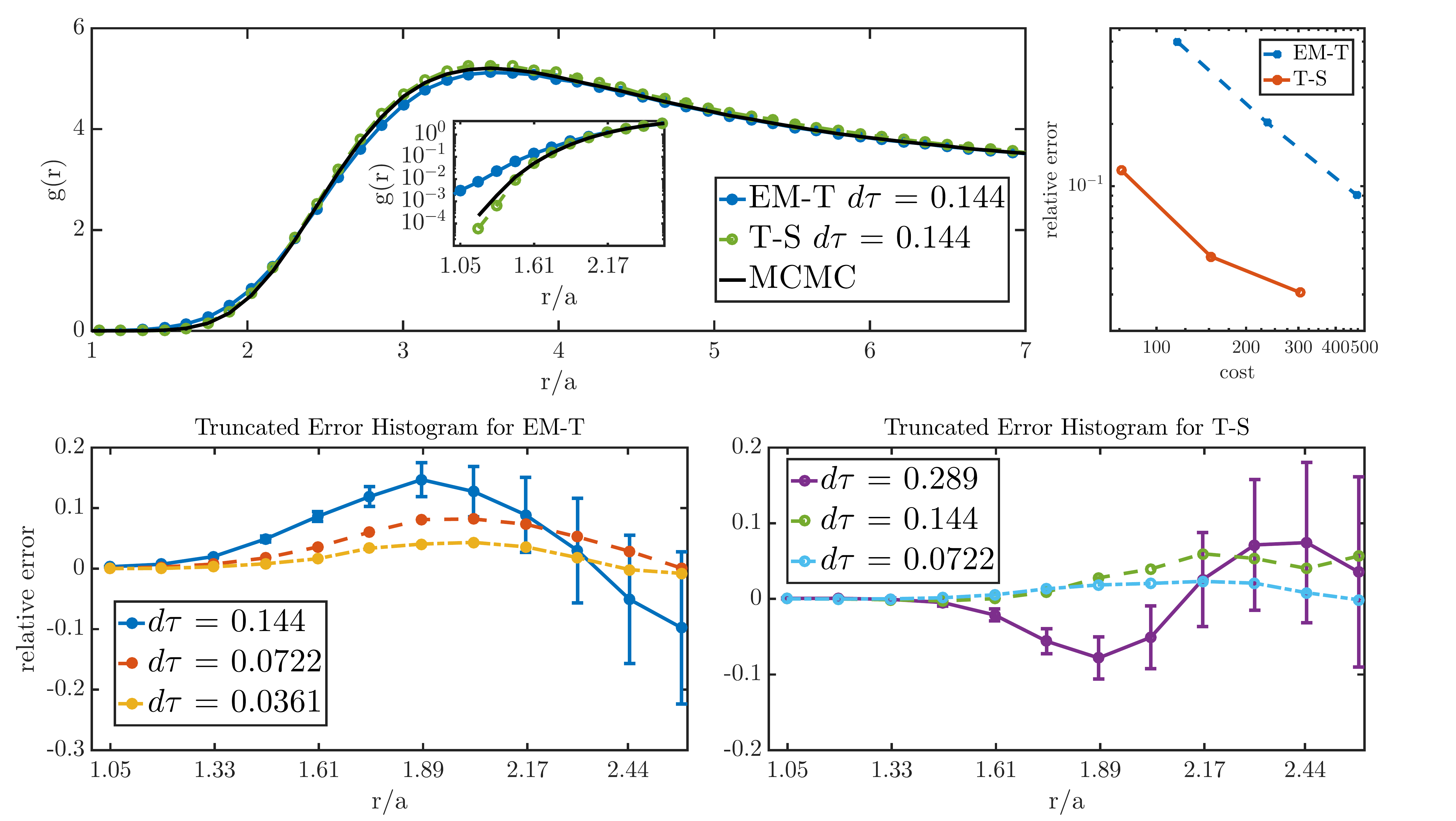}
\caption{\label{fig:gr}Accuracy of the radial distribution function $g(r)$
for a dense suspension of boomerangs (illustrated in the bottom panel
of Fig. \ref{fig:Booms}), for several values of the normalized time
step size $\protect\D{\tau}$. (Upper left panel) Comparison with
the correct $g(r)$ computed using an MCMC method, for $\protect\D{\tau}=0.144$
for the EM-T and T-S schemes. The semi-log scale inset zooms in on
nearly touching boomerangs to reveal a much larger amount of particle
(near) overlaps for the EM-T scheme. (Upper right panel) Cumulative
error as a function of cost per unit time (see main text). (Lower
panels) Error in the radial distribution function $g(r)$ for the
EM-T (left panel) and T-S (right panel) schemes for several different
time step sizes (see legend). Error bars indicate $95\%$ confidence
intervals.}
\end{figure}

The bottom two panels of Fig. \ref{fig:gr} show that the T-S scheme
is notably more accurate than the EM-T scheme for the same value of
the time step $\D{\tau}$. It is also worthy of note that we were
able to run the T-S scheme with fair accuracy using $\D{\tau}=0.29$,
which was seen to be unstable for the EM-T scheme. However, while
the T-S scheme is more accurate, it also requires one more mobility
solve per time step than the EM-T scheme. Therefore, to really determine
which scheme is best for large-scale simulation, we must define a
notion of accuracy and cost and determine which scheme achieves a
given level of accuracy for a smaller computational cost.

We define a cumulative measure of accuracy using a modified $L_{2}$
error of the $g(r)$ histograms relative to reference values computed
with high statistical accuracy using an MCMC algorithm. We account
for statistical errors by considering a weighted $L_{2}$ norm proportional
to the log-likelihood, 
\begin{equation}
\text{Error}\coloneqq\sqrt{{\displaystyle \frac{1}{2}\int_{0}^{R}\left(\frac{g^{\D t}(r)-g_{\text{MCMC}}(r)}{\sigma(r)}\right)^{2}dr}},\label{L2W}
\end{equation}
where $\sigma(r)$ is the standard deviation estimated empirically
using multiple independently-seeded simulations. We take $R=2.5a$
since for $r\gtrsim2.5a$ the error in $g(r)$ is dominated by sampling
(statistical) error for both schemes. 

Since in our specific case the computational cost is dominated by
(dense) multiplications of the blob-blob mobility matrix with a vector,
we define the cost per unit time as the (average) total number of
multiplications by $\Mob$ per time step, divided by $\D{\tau}$.
We observe that when a solver tolerance of $10^{-3}$ is used, the
preconditioned iterative methods to solve the saddle point system,
and to compute $\Mob^{1/2}$, will both converge in 5 iterations most
of the time. Thus, the total number of multiplications per time step
for the T-S scheme is 22 ($3\times5+5+2$ for three mobility solves,
one Lanczos iteration, and one RFD on $\Mob$), while the EM-T scheme
requires 17 ($2\times5+5+2$). 

The upper right panel of Fig. \ref{fig:gr} shows that the T-S scheme
costs less per unit time than the EM-T scheme for any desired accuracy.
We will therefore use the T-S schemes in Section \ref{sec:BrownRoll},
and recommend it for suspensions confined above a no-slip wall. Nevertheless,
the cost of each scheme depends heavily on how expensive it is to
compute the action of $\Mob$ and $\Mob^{\frac{1}{2}}$, and we recommend
repeating the cost-accuracy balance computations reported here for
each specific application/code.

\subsection{\label{sec:BrownRoll}Uniform suspensions of Brownian Rollers}

\begin{figure}[h]
\begin{centering}
\includegraphics[width=0.9\textwidth]{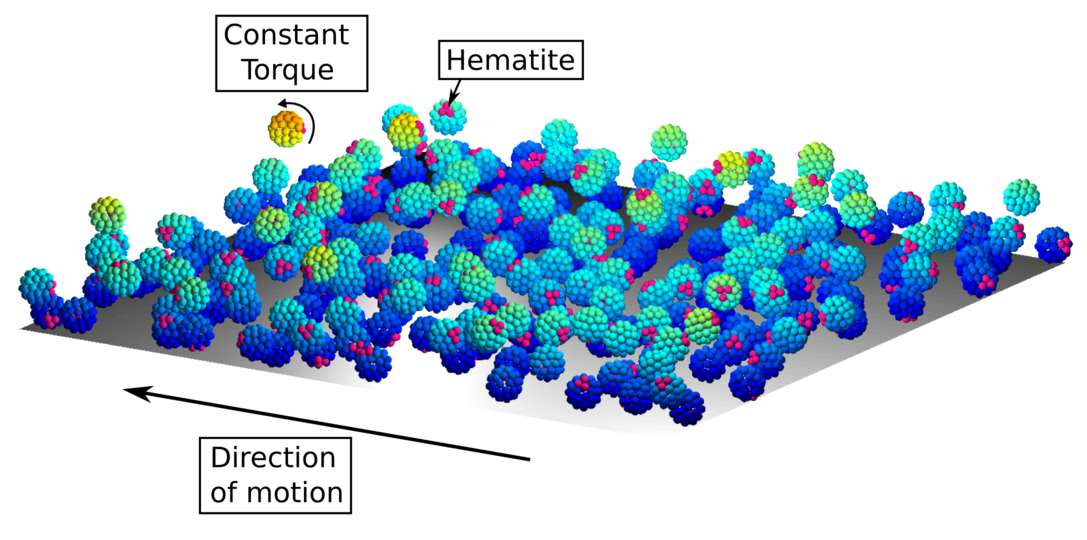}
\par\end{centering}
\centering{}\caption{\label{fig:RollersSnapshot}A snapshot of a steadily-translating uniform
suspension of 256 microrollers, each made up of 42 blobs (colored
by their height above the floor), at planar packing density $\phi=0.4$.
Each particle has an embedded magnet, illustrated as a cluster of
fuchsia blobs. Note that although for constant applied torque the
particle orientation does not enter in the equations of motion for
translation, our algorithm keeps track of the orientation of each
colloid, which can be used to more accurately compute a time-dependent
magnetic torque on the particles if desired.}
\end{figure}

Active suspensions of rotating spherical colloids (microrollers) sedimented
above a bottom wall have been recently investigated using both experiments
and simulations \cite{Rollers_NaturePhys}. The colloids have an embedded
hematite which makes them weakly ferromagnetic and thus easily rotated
by an external magnetic field, as illustrated in Fig. (\ref{fig:RollersSnapshot}).
Because of the presence of a nonzero rotation-translation coupling
due to the bottom wall, micro-rollers translate parallel to the wall.
Collective flow effects dominate the dynamics of many-body suspensions,
and the particles translate much faster at larger densities. For non-uniform
suspensions, shocks were observed to form and destabilize into fingering
instabilities, and deterministic simulations were performed to interrogate
the observations. In \cite{MagneticRollers}, the effects of Brownian
motion were included in the simulations to demonstrate the quantitative
effect that fluctuations have on the the development and progression
of the fingering instability. In particular, it is important to note
that Brownian motion sets the equilibrium gravitational height of
the colloids, and therefore must be included to obtain quantitative
predictions that can directly be compared to experiments. In \cite{NonlocalShocks_Rollers},
the nonlocal nature of the shock front was further elucidated, and
propagation of density waves in a uniform suspension translating parallel
to the wall was investigated using both experiments and simulation.
One of the key parameters that enters in the simplified equations
describing the dynamics of density fluctuations (see Eq. (4) in \cite{NonlocalShocks_Rollers})
around a uniform state is the mean suspension velocity $\bar{V}$.

At higher densities, the mean velocity $\bar{V}$ is dominated by
collective effects and near-field hydrodynamic interactions between
the particles and between the particles and the wall. In all prior
work \cite{MagneticRollers,NonlocalShocks_Rollers}, rollers were
represented using only one blob, and the Rotne-Prager-Blake tensor
was used to add the active translation as a deterministic forcing
term. In \cite{RigidMultiblobs}, it was demonstrated that using more
blobs to discretize spherical particles gives much greater accuracy
for hydrodynamics. We are here able to, for the first time, consistently
and sufficiently accurately resolve hydrodynamics \emph{and} account
for thermal fluctuations, and thus obtain \emph{quantitative} predictions
that can be compared to experiments. Following \cite{MagneticRollers},
we take $\eta=1$ mPa$\cdot$s, the hydrodynamic radius of the particles
$R_{h}=0.656\ \mu$m, excess (buoyant) mass $m_{e}g=1.24\times10^{-14}\frac{\text{kg m}}{\text{s}^{2}}$,
and apply a constant, identical torque on every particle, $\V T=8\pi\eta\omega R_{h}^{3}\widehat{\V y}$,
where we take the angular frequency $\omega=10$Hz. We use the T-S
scheme with $\D t=0.008$s. The particle-particle and particle-wall
interaction potentials are as described in \cite{MagneticRollers}.
We discretize the rollers using 1, 12, or 42 blobs (illustrated in
Fig. (\ref{fig:RollersSnapshot})), following \cite{RigidMultiblobs}.
It is important to note that for $12$ or $42$ blobs per particle
the translation-rotation coupling inducing the active motion is captured
by the multiblob model itself rather than added by hand as it is for
a single blob. After an initial, transient period, we computed individual
particle velocities over intervals of $1/24$s, and collected histograms
of particles' velocities at the steady translating state \footnote{Note that the analysis in \cite{NonlocalShocks_Rollers} shows that
the steady uniform translating state is stable, and that density fluctuations
propagate as waves without growing or shrinking in time.}. Different time intervals to compute the velocity were also explored
but no substantial change was observed.

\begin{figure}[h]
\centering{}\includegraphics[width=1.05\textwidth]{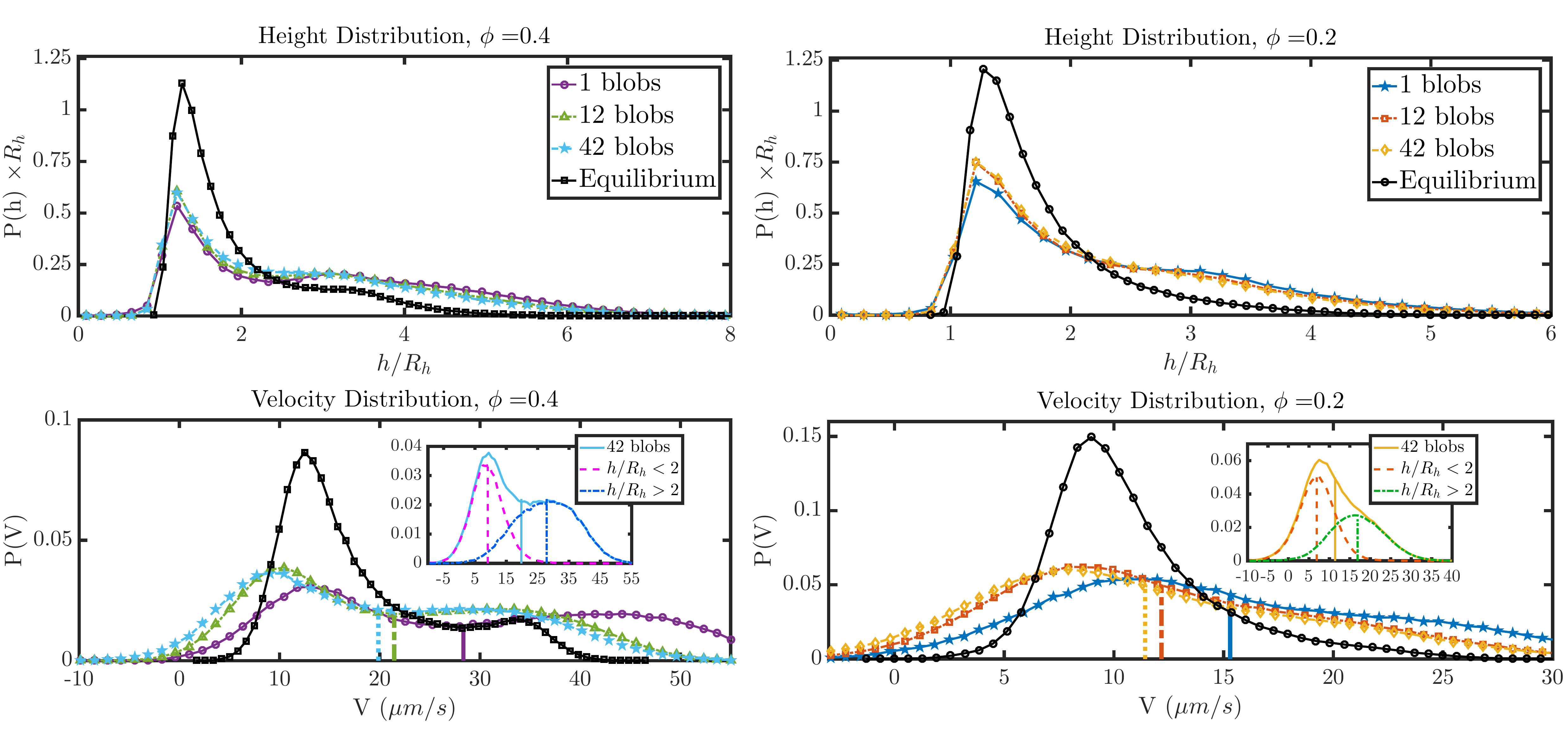}
\caption{\label{fig:rollers}Histograms of the microrollers' heights above
the wall ($P(h)$, top panels) and their velocities ($P(V)$, bottom
panels), for two packing densities in the plane, $\phi=0.2$ (left
panels) and $\phi=0.4$ (right panels), for a (pseudo)periodic active
suspension at steady state. Solid vertical lines demarcate the mean
of the velocity distributions. Curves marked ``equilibrium'' use
particle positions sampled from the equilibrium GB distribution (in
the absence of activity) using an MCMC method. All other curves are
results of dynamic BD simulations using the T-S scheme and either
1, 12 or 42 blobs to resolve each spherical colloid. (Top panels)
Comparison of the height distribution $P(h)$ for $\phi=0.2$ (right)
and $\phi=0.4$ (left), as set by a balance of thermal noise, active
vertical flows and gravity. (Bottom panels) Comparison of the velocity
distribution $P(V)$ for $\phi=0.2$ (right) and $\phi=0.4$ (left).
For the curves marked as equilibrium, velocities were generated by
solving a deterministic mobility problem with particles discretized
by 42 blobs, using configurations sampled by an MCMC method. Insets
show $P(V)$ for the finest resolution split into two groups based
on particle height ($h\lessgtr2R_{h}$), where the normalization factor
for the distributions is based on the fraction of the total number
of particles in the given subgroup.}
\end{figure}

Figure \ref{fig:rollers} shows histograms of the particles' heights
above the wall ($P(h)$, top panel) and their velocities ($P(V)$,
bottom panels), for two packing densities in the plane, $\phi=n\left(\pi R_{h}^{2}\right)=0.2$
(left panels) and $\phi=0.4$ (right panels), where $n$ is the number
density in the plane. We use 256 particles for each case; no significant
change in the results was observed when 1024 particles were used,
confirming that finite size effects are small. The equilibrium Gibbs-Boltzmann
height distributions were computed by using an MCMC method, and are
not affected by the resolution of the multiblobs, which only changes
the (hydro)dynamics of the suspension. To understand the effects on
$P(V)$ due to changes in $P(h)$ caused by the active motion, we
estimate the distribution of particles velocities at a fictitious
``equilibrium'' state by sampling particle positions from the equilibrium
GB distribution using MCMC. We then apply a constant torque $\V T$
on each particle and solve a \emph{deterministic} mobility problem
(using the 42-blob model of the particles) for the particles' velocities.
The bottom panels in Fig. \ref{fig:rollers} show a large mismatch
between these static predictions and the actual dynamics of the particles.
This underscores the importance of explicitly simulating the Brownian
dynamics in this system in order to sample the out-of-equilibrium
steady state distribution, which is quite different from the GB distribution
due to the strong active flows.

The bottom panels in Fig. \ref{fig:rollers} demonstrate that the
more resolved results for $P(V)$ obtained using 12 and 42 blobs closely
match for both packing densities, while the minimally-resolved distributions
using just one blob per particle are fairly dissimilar from the others.
This same mismatch in distribution based on particle resolution is
also seen in the height distributions for both packing densities,
though it is less pronounced. While this certainly indicates that
minimally-resolved simulations are not sufficient make quantitative
predictions, all of the particle resolutions considered produce qualitatively
similar results. More specifically, we estimate the mean velocity
among all particles in $\mu\mbox{m}/s$ for $\phi=0.2$ to be $\bar{V}=15.3$
for 1 blob, $\bar{V}=12.2$ for 12 blobs, $\bar{V}=11.4$ for 42 blobs
per sphere, compared to $\bar{V}=10.9$ predicted by ``equilibrium''
static sampling (with 42 blobs). For $\phi=0.4$, we estimate $\bar{V}=28.4$
for 1 blob, $\bar{V}=21.3$ for 12 blobs, $\bar{V}=19.8$ for 42 blobs
per sphere, compared to $\bar{V}=17.9$ predicted by ``equilibrium''
static sampling. These results also indicate that using as few as
12 blobs per sphere can give sufficiently accurate predictions (with
relative error less than $10\%$) to be quantitatively compared to
experimental measurements. Note that the mean velocity obtained from
the most-resolved computations leads to an estimate of the Péclet
number $\text{Pe}=\left(6\pi\eta\right)R_{h}^{2}\bar{V}/\left(k_{B}T\right)$
of $\text{Pe}\approx22$ for $\phi=0.2$ and $\text{Pe}\approx39$
for $\phi=0.4$. This indicates that the horizontal motion is dominated
by the active flow. However, the Brownian motion is important for
setting the height of the particles above the bottom wall, as can
be seen from the fact that the distribution of heights $P\left(h\right)$
is not changed significantly by the active motion.

All of the simulation results in Fig. \ref{fig:rollers} show bimodal
distributions for both the particles' heights and velocities. Particularly
prominent in the particle velocity distribution for $\phi=0.4$, but
present for $\phi=0.2$ as well, are two peaks indicating the existence
of two distinct populations of ``fast'' and ``slow'' particles.
Close examination of the height distributions reveals a similar bimodality,
and our simulations indicate a strong correlation between particle
height and velocity. We separate particles into two subgroups roughly
corresponding to the two peaks in $P(h)$, and identify the fast particles
as the group corresponding to $h>2R_{h}$, while the remaining particles
are slower, as seen in the inset figures in the lower panels of Fig.
\ref{fig:rollers}. This separation is surprising as we might expect
the opposite given that a single particle will translate faster if
it is placed closer to the wall. This indicates the importance of
collective flows and packing effects in these suspensions. Physically,
the higher packing density causes a relatively dense monolayer of
particles to form around the gravitational height, $h_{G}=R_{h}+k_{B}T/m_{e}g\approx1\mu\mbox{m}$.
The rest of the particles form a sparser and more diffuse (in the
vertical direction) monolayer above the first at height of roughly
$2h_{G}$, and are rapidly advected by the collective flow as they
``slide'' on top of the bottom layer.

The presence of two populations of particles at different heights
and moving at different velocities makes experimental measurements
of $P(V)$ or even $\bar{V}$ more difficult. Namely, at these packing
densities it is not possible to track individual particles to measure
individual particle velocities, and indirect method such as particle
image velocimetry (PIV) are used to estimate $\bar{V}$, which can
lead to bias in the presence of fast and slow particles. Direct comparison
of our computational estimates to experimental measurements is therefore
deferred for future work.

\section{Conclusions}

In this work we designed efficient and robust temporal integrators
for the simulation of many rigid particles suspended in a fluctuating
viscous fluid. Hydrodynamic interactions were computed using a rigid
multiblob model \cite{RigidMultiblobs} of the particles, and here
we proposed a method to generate the Brownian increments of the particles
at a computational cost that is no larger than that of solving a mobility
problem. We demonstrated that the block-diagonal preconditioner used
to solve mobility problems in \cite{RigidMultiblobs} is equally effective
as a preconditioner for the Lanczos algorithm to compute Brownian
increments for the blobs. The stochastic drift term arising from the
configuration-dependent mobility matrix were computed using traction-corrected
or slip-corrected random finite differences. We presented a traction-corrected
Euler-Maruyama scheme (EM-T) (algorithm \ref{alg:traction}), as well
as a slip-corrected Trapezoidal scheme (T-S) (algorithm \ref{alg:sliptrap}).
We have made our python implementation (with PyCUDA acceleration)
of the methods described here available at \url{https://github.com/stochasticHydroTools/RigidMultiblobsWall}.
Both the EM-T and T-S schemes scale linearly in complexity with the
number of rigid particles being simulated if the iterative methods
used to compute deterministic and Brownian blob velocities are based
on fast methods. We confirmed that both schemes correctly reproduce
the equilibrium Gibbs-Boltzmann distribution for sufficiently small
time step sizes, and found the T-S scheme to be notably superior in
accuracy for the same computational effort for large-scale problems.
We used the T-S scheme to study the non-equilibrium dynamics of an
active suspension of microrollers confined above a no-slip bottom
wall, and demonstrated that as few as 12 blobs per sphere gives numerical
errors on the order of 10\% or less, unlike previous simulations of
existing microroller experiments \cite{MagneticRollers,NonlocalShocks_Rollers}.
The use of particle image velocimetry (PIV) in \cite{Rollers_NaturePhys}
to experimentally estimate the average particle velocity can be biased
from the bimodality of the particle velocities, rendering a direct
comparison with experimental results currently unobtainable. In section
\ref{sec:BrownRoll}, we identified two well separated, unimodal,
populations of microrollers demarcated by their heights. This separation
can perhaps be used to design new experimental techniques to accurately
measure the particles' velocities within each population.

There are number of application-dependent parameters that can be tweaked
to improve efficiency. For instance if small particle displacements
are expected over a time step, one can wait several time steps before
recomputing the (Cholesky) factorizations of $\Mob$ that enter in
the block-diagonal preconditioner. Here we used the same relative
error tolerance for all iterative methods, and found a relatively
loose tolerance of $10^{-3}$ to be sufficient. However, one could
use a different solver tolerances in, for example, steps \ref{step:TracStep1},\ref{step:MhalfEM},
and \ref{step:Udef} of algorithm \ref{alg:traction} for the EM-T
scheme. Further, while the EM-T and T-S schemes were found to be optimal
for the applications considered herein, other schemes such as an Adams-Bashforth
variant of the EM-T scheme, or a midpoint variant of the T-S scheme
are straightforward extensions and may prove optimal for other applications
and implementations. In fact, our biggest contribution here is the
development of the traction-corrected and slip-corrected RFDs, which
can be used as tools to construct other schemes. It is important to
realize that showing theoretically that a certain scheme is consistent
in the limit $\D t\rightarrow0$ is \emph{not} sufficient \textendash{}
establishing numerically that the scheme is robust for $\D t$ on
the order of the \emph{diffusive} time scale, as we have done here
for the EM-T and T-S schemes, is crucial. While we do not have a detailed
theoretical understanding of the errors that arise for finite $\D t$,
one important consideration is that RFDs only give the stochastic
drift term in expectation, and it is important to control and reduce
their variance and not just their mean.

The methods presented in this work are fairly general and can be extended
to other geometries and ways of computing hydrodynamic interactions.
In periodic domains, we can use the Positively Split Ewald (PSE) method
\cite{SpectralRPY} to compute deterministic and Brownian blob velocities,
and no change is made in algorithms \ref{alg:traction} or \ref{alg:sliptrap}
to account for this. However, in this case, generating Brownian blob
displacements becomes rather inexpensive compared to a mobility solve,
and it is possible that a midpoint split scheme would become preferable
over the T-S scheme in terms of efficiency-accuracy balance. Note
that the PSE method could be extended to other geometries such as
doubly-periodic domains (e.g., membranes) by building on recently-developed
Spectral Ewald methods \cite{SpectralEwald_DoublyPeriodic,BoundaryIntegral_Wall}.
To the best of our knowledge, there is presently no known method to
compute Brownian blob increments for infinite unbounded domains in
(near) linear time; we relied here explicitly on the the simplicity
of the Rotne-Prager-Blake tensor and the screening by the wall to
handle particles confined in a half-space.

All of the computations performed in this work used a rather coarsely
resolved multiblob model to represent the rigid bodies. In future
work, we will apply the temporal integrators proposed here to more
accurate representations of the geometry and hydrodynamics using the
recently-developed Fluctuating Boundary Integral Method (FBIM) \cite{FBIM}.
Both of the temporal integrators presented here can be used without
modification with FBIM, but a midpoint scheme may be preferable because
of the low-cost of computing Brownian terms compared to solving mobility
problems.

All of the methods presented herein relied extensively on an explicit
representation of $\Mob$, which restricts the choice of domain and
boundary conditions that we may use to those for which an analytical
expressions (and preferably a fast method to compute its action) for
the RPY mobility is available. In \cite{BrownianBlobs} the Stokes
equations are solved explicitly on an Eulerian grid for fully confined
domains such as slit channels, and Immersed Boundary (IB) interpolation
and spreading operators are used to couple the blobs to the fluid
solver. This Green's-function-free or ``explicit-solvent'' (but
still inertia-less) approach implicitly computes the action of $\Mob$
in linear time in the number of fluid grid cells. Some of us demonstrated
in \cite{BrownianBlobs} that the action of $\Mob^{\frac{1}{2}}$
can also be computed using the IB approach at minimal additional cost
by using fluctuating hydrodynamics. In \cite{RigidIBM,RigidMultiblobs},
the IB approach was extended to rigid bodies (multiblobs), but without
accounting for Brownian motion. The schemes presented herein can,
in principle, be used with only minor modification with the rigid-body
IB method to simulate Brownian motion of rigid particles in fully
confined domains, when explicit representation of $\Mob$ is not available.
However, the efficiency of the methods used in this work hinged on
the action of $\Mob$ being computed rapidly, and hence the temporal
integrators should be modified to account for the comparatively expensive
explicit-solvent IB solvers introduced in \cite{RigidMultiblobs}.
Efficient simulation of rigid, Brownian particles in general confined
domains will be the subject of future work.
\begin{acknowledgments}
We are grateful to Blaise Delmotte for his help with simulations of
active roller suspensions, and to Michelle Driscoll and Paul Chaikin
for numerous discussions about experiments on microrollers. This work
was supported in part by the National Science Foundation under collaborative
award DMS-1418706 and by DMS\textendash 1418672, and by the U.S. Department
of Energy Office of Science, Office of Advanced Scientific Computing
Research, Applied Mathematics program under award DE-SC0008271. We
thank the NVIDIA Academic Partnership program for providing GPU hardware
for performing the simulations reported here.
\end{acknowledgments}

\newpage{}

\section*{Appendix}

\appendix
%dummy comment inserted by tex2lyx to ensure that this paragraph is not empty

\section{\label{add:EMrfd}Euler-Maruyama Scheme}

In section \ref{sec:RFD} we noted that a simple means of computing
an RFD on $\sM B\equiv\N$, is to use $\Delta\V P=\Delta\V Q=\sqrt{k_{B}T}\,\V W$
in equation (\ref{eq:RFDB}), where $\V W$ is a vector drawn from
the standard normal distribution. However, incrementing the translational
and rotational components of the configuration $\V Q$ by the same
quantity may cause translation and rotation of a body by very different
magnitudes. This, in turn, may result in large variance of the quantity
$\partial_{\V Q}\N\colon\left[\V W\V W^{T}\right]$ (see (\ref{eq:RFDB})),
and hence slow the convergence to $\partial_{\V Q}\cdot\N$ in expectation.
To remedy this, we simply ensure that a given body is being randomly
translated and rotated by the same amount. 

Specifically, we choose a length scale, $L_{p}$, to represent the
size of body $p$, and compute the random displacement,
\begin{equation}
\Delta\V Q_{p}=\begin{bmatrix}L_{p}\V W_{p}^{f}\\
\V W_{p}^{\tau}
\end{bmatrix}
\end{equation}
as well as the random forces and torques,
\begin{equation}
\V W_{p}^{FT}=k_{B}T\begin{bmatrix}L_{p}^{-1}\V W_{p}^{f}\\
\V W_{p}^{\tau}
\end{bmatrix}
\end{equation}
where $\V W_{p}^{f},\V W_{p}^{\tau}$ are both three dimensional standard
Gaussian random vectors, and we form the composite vectors $\V W^{FT}=\left[\V W_{p}^{FT}\right]$
and $\Delta\V Q=\left[\Delta\V Q_{p}\right]$. The drift can then
be computed using a simple RFD, 
\begin{equation}
\frac{1}{\delta}\av{\left\{ \N\left(\V Q+\frac{\delta}{2}\Delta\V Q\right)-\N\left(\V Q-\frac{\delta}{2}\Delta\V Q\right)\right\} \V W^{FT}}=k_{B}T\ \partial_{\V Q}\cdot\N+O(\delta^{2}),\label{eq:RFDL}
\end{equation}
which respects the physical units of the problem and minimizes the
variance of the approximation. A one-sided difference approximation
can be defined analogously.

Algorithm \ref{alg:RFDN} summarizes a scalable implementation of
the Euler-Maruyama RFD scheme from \cite{BrownianMultiBlobs} to solve
(\ref{Langevin}), using a random finite difference similar to (\ref{eq:RFDL})
to compute the drift. Implementation of this scheme requires three
solutions of the saddle point system (\ref{eq:saddle}) and one Lanczos
application per timestep.%

\begin{algorithm}[h]
\caption{\label{alg:RFDN} Euler-Maruyama-RFD scheme}
\begin{enumerate}
\item Compute RFD terms:

\begin{enumerate}
\item Form $\V W^{FT}=\left[\V W_{p}^{FT}\right],$ where 
\[
\V W_{p}^{FT}=k_{B}T\begin{bmatrix}L_{p}^{-1}\V W_{p}^{f}\\
\V W_{p}^{\tau}
\end{bmatrix}.
\]
\item Displace the particles by small random amounts: 
\[
\V Q_{p}^{\pm}=\V Q_{p}^{n}\pm\frac{\delta}{2}\begin{bmatrix}L_{p}\V W_{p}^{f}\\
\V W_{p}^{\tau}
\end{bmatrix}.
\]
\item \label{enu:EM_RFD_solve}Solve \textit{two} mobility problems for
$\V U^{+},\V U^{-}$:
\[
\begin{bmatrix}\Mob^{\pm} & -\K^{\pm}\\
-\left(\K^{T}\right)^{\pm} & \V 0
\end{bmatrix}\begin{bmatrix}{\V{\lambda}}^{\pm}\\
\V U^{\pm}
\end{bmatrix}=\begin{bmatrix}\V 0\\
-\V W^{FT}
\end{bmatrix}.
\]
\end{enumerate}
\item Compute $\left(\Mob^{1/2}\right)^{n}\V W^{n}$ using a preconditioned
Lancoz method or PSE.
\item Solve mobility problem:
\[
\begin{bmatrix}\Mob^{n} & -\K^{n}\\
-\left(\K^{T}\right)^{n} & \V 0
\end{bmatrix}\begin{bmatrix}{\V{\lambda}}^{n}\\
\V U^{n}
\end{bmatrix}=\begin{bmatrix}-\sqrt{\frac{2k_{B}T}{\Delta t}}\left(\Mob^{1/2}\right)^{n}\V W^{n}\\
-\V F^{n}
\end{bmatrix}.
\]
\item Update configuration:
\[
\V Q^{n+1}=\V Q^{n}+\Delta t\left\{ \V U^{n}+\frac{1}{\delta}(\V U^{+}-\V U^{-})\right\} .
\]
\end{enumerate}
\end{algorithm}

\section{\label{Add:Proof}Proofs of Consistency of Temporal Integrators}

In this appendix we demonstrate that the two temporal integrators
presented in this work generate the required stochastic drift terms.
In all of the calculations of this section, we will use index notation
with the convention of summing over repeated indices. We will also
use the convention that superscripted indices correspond to a matrix
inverse, i.e $\left[\sM A^{-1}\right]_{ij}\equiv\sM A^{ij}$. To avoid
possible confusion with our superscript notation for the time level
at which an operator is evaluated, we will assume that all terms and
operators are evaluated at the base configuration $\V Q\equiv\V Q^{n}$
unless otherwise noted. We denote partial derivatives with the shorthand
notation $\partial_{k}\equiv\partial/\partial Q_{k}$.

\subsection{\label{Add:Quat}Overdamped Langevin Equation using Quaternions}

In principle, any means of representing orientation could be used
with the techniques detailed in this work. In previous work \cite{BrownianMultiBlobs},
unit quaternions were found to be a particularly favorable choice,
and we will use normalized quaternions to represent the orientation
$\V{\theta}$ hereafter. Here we briefly review key notation and results
regarding quaternions; details can be found in \cite{BrownianMultiBlobs}.
A normalized quaternion is a vector $\V{\theta}=\left[s,\V p\right]\in\Set R^{4}$
such that $\norm{\V{\theta}}_{2}=1$. We define an orientation dependent
$4\times3$ ``projection'' matrix $\V{\Psi}$ as 
\begin{equation}
\rot\left(\V{\theta}\right)=\frac{1}{2}\begin{bmatrix}-\V p^{T}\\
s\M I-\M P
\end{bmatrix},
\end{equation}
where $\M P$ is the cross product matrix defined by $\M P\V x=\V p\times\V x$.
For a configuration, $\V Q=\left[\V q,\V{\theta}\right]$ and a rigid
body velocity $\V U=\left[\V u,\V{\omega}\right]$, we introduce a
matrix 
\begin{equation}
\Rot=\begin{bmatrix}\M I & \M 0\\
\M 0 & \rot
\end{bmatrix},
\end{equation}
so that we may write $d\V Q/dt=\M{\Rot}\V U$. In \cite{BrownianMultiBlobs}
some of us showed that the overdamped Langevin Ito equation (\ref{Langevin})
may be written using unit quaternions as
\begin{eqnarray}
\frac{d\V Q_{l}}{dt} & = & \Rot_{li}\N_{ij}\V F_{j}+k_{B}T\ \left(\Rot_{li}\left\{ \partial_{k}\N_{ij}\right\} \Rot_{kj}+\left\{ \partial_{k}\Rot_{li}\right\} \N_{ij}\Rot_{kj}\right)+\sqrt{2k_{B}T}\;\Rot_{li}\N_{it}^{1/2}\sM W_{t}\label{eq:quatLangevin}\\
 & = & \Rot_{li}\circ\left(\N_{ij}\V F_{j}+k_{B}T\ \left\{ \partial_{k}\N_{ij}\right\} \Rot_{kj}+\sqrt{2k_{B}T}\;\N_{it}^{1/2}\sM W_{t}\right)\nonumber \\
 & = & \Rot_{li}\circ\V U_{i},\nonumber 
\end{eqnarray}
where $\circ$ denotes the Stratonovich product. As shown in \cite{BrownianMultiBlobs},
the forces and torques of the system may be written as $\V F_{j}=-\M{\Rot}_{kj}\partial_{k}U,$
where $U\left(\V Q\right)$ is the conservative potential. This, combined
with the fact that $\partial_{k}\Rot_{kj}=0$, allows us to recognize
(\ref{eq:quatLangevin}) as having the same form as equation (\ref{Langevin})
but with $\N$ replaced by $\Rot\N\Rot^{T}$. 

In order to discretize (\ref{eq:quatLangevin}), we write an expansion
in $\Delta t$ for the procedure which rotates a quaternion $\V{\theta}$
by an angular displacement $\V{\omega}\Delta t$ as 
\begin{equation}
\text{Rotate}\left(\V{\theta}_{k},\V{\omega}_{k}\Delta t\right)\approx\V{\theta}_{k}+\Delta t\rot_{kj}\V{\omega}_{j}-\frac{\Delta t^{2}}{8}\norm{\V{\omega}}_{2}^{2}\V{\theta}_{k}.\label{eq:rotate}
\end{equation}
The second order term in (\ref{eq:rotate}) is responsible for the
fact that that a simple (inconsistent) Euler-Maruyama scheme without
an RFD, using a velocity 
\begin{equation}
\V U_{i}=\N_{ij}\V F_{j}+\sqrt{\frac{2k_{B}T}{\Delta t}}\N_{it}^{1/2}\V W_{t}^{n}\label{eq:EM_U}
\end{equation}
to update the configurations to the next time step, will generate
the drift term $\left(k_{B}T\right)\left\{ \partial_{k}\Rot_{li}\right\} \N_{ij}\Rot_{kj}$
in (\ref{eq:quatLangevin}) to leading order in $\Delta t$ \cite{BrownianMultiBlobs}.
This is to say that the rotate procedure (\ref{eq:rotate}) naturally
captures the Stratonovich product in the second line of (\ref{eq:quatLangevin}).
A one-step numerical scheme is first-order weakly accurate if the
first three moments of the numerical update are correct to $O(\D t)$
\cite{MilsteinSDEBook}; the third moment in our schemes is easily
seen to be at least $O(\D t^{3/2})$. Hence, to show that the schemes
introduced in this work are weakly first order in solving (\ref{eq:quatLangevin}),
we will show that velocity used to update the configurations to the
next time step is of the form
\begin{equation}
\V U_{i}=\N_{ij}\V F_{j}+\sqrt{\frac{2k_{B}T}{\Delta t}}\N_{it}^{1/2}\V W_{t}^{n}+k_{B}T\ \left\{ \partial_{k}\N_{ij}\right\} \Rot_{kj}+\mathcal{R}\left(\D t,\,\D t^{1/2}\right),\label{eq:EM_RFD_U}
\end{equation}
where the notation $\mathcal{R}\left(a,b^{1/2}\right)$ denotes a
Gaussian random error term with mean $O\left(a\right)$, and variance
$O\left(b\right)$. Notice that the leading order term in (\ref{eq:EM_RFD_U}),
the Brownian velocity $\sqrt{\frac{2k_{B}T}{\Delta t}}\N_{it}^{1/2}\V W_{t}^{n}$,
controls the second moment of the velocity update. This term is easily
identified in the velocity update produced by the schemes introduced
in this work. The main difficulty is showing that both schemes generate
in expectation value the required drift term $k_{B}T\ \left\{ \partial_{k}\N_{ij}\right\} \Rot_{kj}$
for sufficiently small $\D t$. For this, we will use the chain rule
to expand the derivative,
\begin{align}
\partial_{k}\N_{ij}= & -\N_{im}\left\{ \partial_{k}\N^{ml}\right\} \N_{lj}=-\N_{im}\left(\partial_{k}\left\{ \K_{pm}\Mob^{pq}\K_{ql}\right\} \right)\N_{lj}\nonumber \\
= & -\N_{im}\left\{ \partial_{k}\K_{pm}\right\} \Mob^{pq}\K_{ql}\N_{lj}-\N_{im}\K_{pm}\left\{ \partial_{k}\Mob^{pq}\right\} \K_{ql}\N_{lj}-\N_{im}\K_{pm}\Mob^{pq}\left\{ \partial_{k}\K_{ql}\right\} \N_{lj}\nonumber \\
= & -\N_{im}\left\{ \partial_{k}\K_{sm}\right\} \Mob^{sq}\K_{ql}\N_{lj}\label{eq:DNindex}\\
 & +\N_{im}\K_{pm}\Mob^{pr}\left\{ \partial_{k}\Mob_{rs}\right\} \Mob^{sq}\K_{ql}\N_{lj}\nonumber \\
 & -\N_{im}\K_{pm}\Mob^{pr}\left\{ \partial_{k}\K_{rl}\right\} \N_{lj}.\nonumber 
\end{align}
Note that the expression for $\left(\partial_{\V Q}\cdot\N\right)_{i}$
can be obtained by replacing every instance of the index $k$ with
$j$ in the above. For clarity but without loss of generality, we
take $\V F=\V 0$ hereafter. Note that a nonzero value of $\V F$
may affect the variance of a velocity update and thus the accuracy
of the scheme for finite values of $\D t$. Still, the weak \textit{order}
of accuracy of the schemes considered will be maintained for nonzero
$\V F$.

\subsection{\label{Add:tractionDense}Traction-Corrected Euler-Maruyama Scheme}

It is immediate from the discussion in section \ref{sec:Nhalf}, that
the Brownian velocities produced by Algorithm \ref{alg:traction}
are $\sqrt{\frac{2k_{B}T}{\Delta t}}\N_{ij}^{1/2}\V W_{j}^{n}$ to
leading order in $\D t$, and hence produce the correct second moment.
We will now show that the velocity produced in step \ref{step:Udef}
of Algorithm \ref{alg:traction} is such that 
\begin{equation}
\av{\V U_{i}^{n}}=k_{B}T\ \left\{ \partial_{k}\N_{ij}\right\} \Rot_{kj}+O\left(\delta^{2}\right).\label{eq:EM_drift_desired}
\end{equation}
First, we may write step \ref{step:Udef} as
\begin{align*}
\V U_{i}^{n} & =-\N_{im}\V D_{m}^{F}+\N_{im}\K_{pm}\Mob^{pr}\left(\V D_{r}^{S}+\sqrt{\frac{2k_{B}T}{\Delta t}}\Mob_{rj}^{1/2}\V W_{j}^{n}\right)\\
 & =-\N_{im}\V D_{m}^{F}+\N_{im}\K_{pm}\Mob^{pr}\V D_{r}^{S}+\sqrt{\frac{2k_{B}T}{\Delta t}}\N_{ij}^{1/2}\V W_{j}^{n}
\end{align*}
and thus
\begin{equation}
\av{\V U_{i}^{n}}=-\N_{im}\av{\V D_{m}^{F}}+\N_{im}\K_{pm}\Mob^{pr}\av{\V D_{r}^{S}}.\label{eq:meanU}
\end{equation}
We can expand the quantities $\V D_{m}^{F}$ and $\V D_{r}^{S}$,
defined in step \ref{step:DFDS}, using the definition of $\V{\lambda}^{\text{RFD}}$
from step \ref{step:ULam},
\begin{align*}
\V U_{l}^{\text{RFD}} & =\N_{lj}{\V W}_{j}^{FT}\\
\V{\lambda}_{s}^{\text{RFD}} & =\Mob^{sq}\K_{ql}\V U_{l}^{\text{RFD}}=\Mob^{sq}\K_{ql}\N_{lj}{\V W}_{j}^{FT}.
\end{align*}

Beginning with $\V D_{m}^{F}$, 
\begin{align}
\av{\V D_{m}^{F}} & =\frac{1}{\delta}\av{\left\{ \K_{sm}\left(\V Q^{+}\right)-\K_{sm}\left(\V Q^{-}\right)\right\} \Mob^{sq}\K_{ql}\N_{lj}{\V W}_{j}^{FT}}\label{eq:lambdaUse}\\
 & =\left\{ \partial_{k}\K_{sm}\right\} \Mob^{sq}\K_{ql}\N_{lj}\av{\Delta\V Q_{k}\ {\V W}_{j}^{FT}}+O(\delta^{2}).\label{eq:DFtrp}
\end{align}
Using the expansion for the rotation procedure from (\ref{eq:rotate}),
we may write $\Delta\V Q_{k}=\left[\D{\V q}_{k},\D{\theta}_{k}\right]$
as 
\begin{align*}
\D{\V q_{k}} & =\frac{\delta}{2}L_{p}\V W_{k}^{f}\\
\D{\V{\theta}_{k}} & =\frac{\delta}{2}\rot_{kl}\V W_{l}^{\tau}+\mathcal{R}\left(\delta^{2},\delta^{2}\right),
\end{align*}
from which it is straightforward to verify that 
\begin{equation}
\av{\Delta\V Q_{k}\ {\V W}_{j}^{FT}}=k_{B}T\ \Rot_{kj}+O(\delta^{2}).\label{eq:DQW}
\end{equation}
Using equations (\ref{eq:DFtrp}) and (\ref{eq:DQW}), we see that
the terms produced by $\V D^{F}$ give one part of the required stochastic
drift term, 
\begin{equation}
\av{\V D_{m}^{F}}=k_{B}T\ \left\{ \partial_{k}\K_{sm}\right\} \Mob^{sq}\K_{ql}\N_{lj}\Rot_{kj}+O(\delta^{2}).\label{eq:DFavg}
\end{equation}

Now, for the contributions from terms involving $\V D_{r}^{S}$, we
have 
\begin{align}
\av{\V D_{r}^{S}}= & \frac{1}{\delta}\av{\left\{ \Mob_{rs}\left(\V Q^{+}\right)-\Mob_{rs}\left(\V Q^{-}\right)\right\} \Mob^{sq}\K_{ql}\N_{lj}{\V W}_{j}^{FT}}\nonumber \\
 & -\frac{1}{\delta}\av{\left\{ \K_{rl}\left(\V Q^{+}\right)-\K_{rl}\left(\V Q^{-}\right)\right\} \N_{lj}{\V W}_{j}^{FT}}\nonumber \\
= & \left(\left\{ \partial_{k}\Mob_{rs}\right\} \Mob^{sq}\K_{ql}-\left\{ \partial_{k}\K_{rl}\right\} \right)\N_{lj}\av{\Delta\V Q_{k}\ {\V W}_{j}^{FT}}+O(\delta^{2}),\label{eq:DStrp}
\end{align}
and hence, using equations (\ref{eq:DStrp}) and (\ref{eq:DQW}),
we see that from $\M D^{S}$ we get two more pieces of the required
stochastic drift term, 
\begin{align}
\av{\V D_{r}^{S}} & =k_{B}T\ \left(\left\{ \partial_{k}\Mob_{rs}\right\} \Mob^{sq}\K_{ql}-\left\{ \partial_{k}\K_{rl}\right\} \right)\N_{lj}\Rot_{kj}+O(\delta^{2}).\label{eq:DSavg}
\end{align}
Plugging the results from equations (\ref{eq:DFavg}) and (\ref{eq:DSavg})
into equation (\ref{eq:meanU}), and using (\ref{eq:DNindex}), gives
\begin{align}
\av{\V U_{i}^{n}}= & -\N_{im}\av{\V D_{m}^{F}}+\N_{im}\K_{pm}\Mob^{pr}\av{\V D_{r}^{S}}+O(\delta^{2})\\
= & -k_{B}T\ (\N_{im}\left\{ \partial_{k}\K_{sm}\right\} \Mob^{sq}\K_{ql}\N_{lj}\Rot_{kj}\nonumber \\
 & +\N\K_{pm}\Mob^{pr}\left\{ \partial_{k}\Mob_{rs}\right\} \Mob^{sq}\K_{ql}\N_{lj}\Rot_{kj}\nonumber \\
 & -\N\K_{pm}\Mob^{pr}\left\{ \partial_{k}\K_{rl}\right\} \N_{lj}\Rot_{kj})+O(\delta^{2})\nonumber \\
= & k_{B}T\ \left\{ \partial_{k}\N_{ij}\right\} \Rot_{kj}+O(\delta^{2}),\nonumber 
\end{align}
where the last equality comes directly from equation (\ref{eq:DNindex}).
This is the desired result (\ref{eq:EM_drift_desired}).

\subsection{\label{Add:SlipDense}Slip-Corrected Trapezoidal Scheme}

We must show that the velocity used to update the position in the
corrector step in Algorithm \ref{alg:sliptrap},
\begin{equation}
\V U_{i}^{\left(c\right)}=\frac{1}{2}\left(\widetilde{\V U}_{i}+\V U_{i}^{n}\right),
\end{equation}
satisfies (\ref{eq:EM_RFD_U}) in law to leading order. We first show
that the predictor-corrector update already gives part of the required
drift term. For this we set $\V D^{F}=\V D^{S}=\V 0$ (recall that
we take $\V F=0$ for simplicity), and show that to leading order
in $\Delta t$, the velocity update in step \ref{step:unun1} of Algorithm
\ref{alg:sliptrap} is such that \footnote{Note that this quantity gives the drift term produced by equation
(\ref{eq:IncTrap2}) in section \ref{sec:Slip}.} 
\begin{equation}
\av{\V U_{i}^{\left(c\right)}}=-k_{B}T\ \N_{im}\K_{pm}\Mob^{pr}\left\{ \partial_{k}\K_{rl}\right\} \N_{lj}\Rot_{kj}+O\left(\D t\right).\label{eq:IncSTdrift}
\end{equation}
To show this we first write steps \ref{step:unT} and \ref{step:unp1T}
as 
\begin{align}
\V U_{j}^{n} & =\sqrt{\frac{2k_{B}T}{\Delta t}}\ \N_{jl}\K_{ql}\Mob^{qs}\Mob_{su}^{1/2}\ \V W_{u}^{n},\label{eq:UnSTproof}\\
\widetilde{\V U}_{i} & =\sqrt{\frac{2k_{B}T}{\Delta t}}\ \aN_{im}\aK_{pm}\aMob^{pr}\Mob_{rt}^{1/2}\V W_{t}^{n},
\end{align}
and write the velocity update of step \ref{step:unun1} as 
\begin{align}
\sqrt{\frac{2\Delta t}{k_{B}T}}\,\V U_{i}^{\left(c\right)}= & \N_{im}\K_{pm}\Mob^{pr}\Mob_{rt}^{1/2}\V W_{t}^{n}+\aN_{im}\aK_{pm}\aMob^{pr}\Mob_{rt}^{1/2}\V W_{t}^{n}\nonumber \\
= & \N_{im}\K_{pm}\Mob^{pr}\Mob_{rt}^{1/2}\V W_{t}^{n}\nonumber \\
 & +\left(\N_{im}\K_{pm}\Mob^{pr}+\partial_{k}\left\{ \N_{im}\K_{pm}\Mob^{pr}\right\} \ \Delta\widetilde{\V Q}_{k}+O\left(\Delta\widetilde{\V Q}^{2}\right)\right)\Mob_{rt}^{1/2}\V W_{t}^{n}\label{eq:STtimesplit}\\
= & 2\N_{im}\K_{pm}\Mob^{pr}\Mob_{rt}^{1/2}\V W_{t}^{n}+\partial_{k}\left\{ \N_{im}\K_{pm}\Mob^{pr}\right\} \Mob_{rt}^{1/2}\ \Delta\widetilde{\V Q}_{k}\V W_{t}^{n}+\mathcal{R}\left(\D t^{3/2},\D t\right)\nonumber \\
= & 2\N_{it}^{1/2}\V W_{t}^{n}+\partial_{k}\left\{ \N_{im}\K_{pm}\Mob^{pr}\right\} \Mob_{rt}^{1/2}\ \Delta\widetilde{\V Q}_{k}\V W_{t}^{n}+\mathcal{R}\left(\D t^{3/2},\D t\right).\label{eq:STlemFinal}
\end{align}
In equation (\ref{eq:STtimesplit}), we have used the fact that all
operators with tilde are evaluated at $\widetilde{\V Q}$, with $\Delta\widetilde{\V Q}=\widetilde{\V Q}-\V Q^{n}$. 

All that remains is to compute, 
\begin{align}
\av{\V U_{i}^{\left(c\right)}}= & \sqrt{\frac{k_{B}T}{2\Delta t}}\ \partial_{k}\left\{ \N_{im}\K_{pm}\Mob^{pr}\right\} \Mob_{rt}^{1/2}\av{\Delta\widetilde{\V Q}_{k}\V W_{t}^{n}}+O\left(\D t\right).\label{eq:meanUU}
\end{align}
Using the expansion for the rotate procedure (\ref{eq:rotate}), we
may write
\begin{align*}
\D{\tilde{\V Q}_{k}} & =\Delta t\Rot_{kj}\V U_{j}^{n}+\mathcal{R}\left(\D t^{2},\D t^{2}\right),
\end{align*}
which combined with (\ref{eq:UnSTproof}) gives 
\begin{align}
\av{\Delta\widetilde{\V Q}_{k}\V W_{t}^{n}}= & \sqrt{2\Delta tk_{B}T}\ \Rot_{kj}\N_{jl}\K_{ql}\Mob^{qs}\Mob_{su}^{1/2}\av{\V W_{u}^{n}\V W_{t}^{n}}+O(\D t^{2})\label{eq:DQtil}\\
= & \sqrt{2\Delta tk_{B}T}\ \Rot_{kj}\N_{jl}\K_{ql}\Mob^{qs}\Mob_{st}^{1/2}+O(\D t^{2}),\nonumber 
\end{align}
where we used $\av{\V W_{u}^{n}\V W_{t}^{n}}=\delta_{tu}$. Hence,
equation (\ref{eq:meanUU}) becomes, 
\begin{align}
\av{\V U_{i}^{\left(c\right)}}= & k_{B}T\ \partial_{k}\left\{ \N_{im}\K_{pm}\Mob^{pr}\right\} \Mob_{rt}^{1/2}\Mob_{ts}^{1/2}\Mob^{sq}\K_{ql}\N_{lj}\Rot_{kj}+O\left(\D t\right)\\
= & k_{B}T\ \partial_{k}\left\{ \N_{im}\K_{pm}\Mob^{pr}\right\} \Mob_{rs}\Mob^{sq}\K_{ql}\N_{lj}\Rot_{kj}+O\left(\D t\right)\nonumber \\
= & k_{B}T\ \partial_{k}\left\{ \N_{im}\K_{pm}\Mob^{pr}\right\} \K_{rl}\N_{lj}\Rot_{kj}+O\left(\D t\right)\nonumber 
\end{align}
where we used that $\Mob_{rs}\Mob^{sq}=\delta_{rq}$. After expanding
$\partial_{k}\left\{ \N_{im}\K_{pm}\Mob^{pr}\right\} $ using the
chain rule and recalling that $\K_{pm}\Mob^{pr}\K_{rl}=\N^{ml}$,
we get
\begin{align}
\av{\V U_{i}^{\left(c\right)}}= & k_{B}T\ \big(\left\{ \partial_{k}\N_{im}\right\} \Rot_{km}\label{eq:DN3term-1-1}\\
 & \qquad+\N_{im}\left\{ \partial_{k}\K_{pm}\right\} \Mob^{pr}\K_{rl}\N_{lj}\Rot_{kj}\nonumber \\
 & \qquad+\N_{im}\K_{pm}\left\{ \partial_{k}\Mob^{pr}\right\} \K_{rl}\N_{lj}\Rot_{kj}\big)+O\left(\D t\right)\nonumber \\
= & -k_{B}T\ \N_{im}\K_{pm}\Mob^{pr}\left\{ \partial_{k}\K_{rl}\right\} \N_{lj}\Rot_{kj}+O\left(\D t\right).\label{eq:DKlast-1-1}
\end{align}
where we have used (\ref{eq:DNindex}). This gives the desired result
(\ref{eq:IncSTdrift}). 

We now include the contribution from nonzero $\V D^{F}$ and $\V D^{S}$
to show that $\V U_{i}^{\left(c\right)}$ satisfies equation (\ref{eq:EM_RFD_U})
as desired. Including the contributions from $\V D^{F}$and $\V D^{S}$,
we may write the velocity update of step \ref{step:unun1} as,
\begin{align}
\V U_{i}^{\left(c\right)}= & -\aN_{im}\V D_{m}^{F}+\aN_{im}\aK_{pm}\aMob^{pr}\V D_{r}^{S}\nonumber \\
 & +\sqrt{\frac{k_{B}T}{2\Delta t}}\N_{im}\K_{pm}\Mob^{pr}\Mob_{rt}^{1/2}\V W_{t}^{n}+\sqrt{\frac{k_{B}T}{2\Delta t}}\aN_{im}\aK_{pm}\aMob^{pr}\Mob_{rt}^{1/2}\V W_{t}^{n}.
\end{align}
Using equations (\ref{eq:STlemFinal}) and (\ref{eq:DQtil}), we may
write this as 
\begin{align}
\V U_{i}^{\left(c\right)}= & -\aN_{im}\V D_{m}^{F}+\aN_{im}\aK_{pm}\aMob^{pr}\V D_{r}^{S}+\sqrt{\frac{2k_{B}T}{\Delta t}}\N_{it}^{1/2}\V W_{t}^{n}\label{eq:slipTEMP}\\
 & +\partial_{k}\left\{ \N_{im}\K_{pm}\Mob^{pr}\right\} \Mob_{rt}^{1/2}\left(\Rot_{kj}\N_{jl}\K_{ql}\Mob^{qs}\Mob_{su}^{1/2}\V W_{u}^{n}\V W_{t}^{n}\right)+\mathcal{R}\left(\D t,\D t^{1/2}\right).\nonumber 
\end{align}
In this form, it is easy to see that the scheme produces the correct
Brownian velocity, $\sqrt{\frac{2k_{B}T}{\Delta t}}\N_{it}^{1/2}\V W_{t}^{n}$,
and all that is left to verify is the first moment. Equation (\ref{eq:IncSTdrift})
allows us to immediately write the mean of (\ref{eq:slipTEMP}) as

\begin{align}
\av{\V U_{i}^{\left(c\right)}}= & \av{-\aN_{im}\V D_{m}^{F}+\aN_{im}\aK_{pm}\aMob^{pr}\V D_{r}^{S}}\label{eq:slipUUfinal}\\
 & -k_{B}T\ \N_{im}\K_{pm}\Mob^{pr}\left\{ \partial_{k}\K_{rl}\right\} \N_{lj}\Rot_{kj}+O\left(\D t\right).\nonumber 
\end{align}

We now examine the terms involving $\V D_{m}^{F}$ and $\V D_{r}^{S}$,
separately. First, we note that using the expanded rotate procedure
(\ref{eq:rotate}), steps \ref{step:QpmT} and \ref{step:UrfdT} in
algorithm \ref{alg:sliptrap} give
\begin{equation}
\V Q_{k}^{\pm}-\V Q_{k}=\pm\delta\ \Rot_{kj}\Delta\V Q_{k}^{\text{RFD}}+\mathcal{R}\left(\delta^{2},\delta^{2}\right)=\pm\delta\ \Rot_{kj}\N_{jl}\K_{lq}\Mob^{qs}\slipW_{s}^{D}+\mathcal{R}\left(\delta^{2},\delta^{2}\right),\label{eq:DeltaQref}
\end{equation}
Using the definition of $\V D_{m}^{F}$ from step \ref{step:DFDS2}
of Algorithm \ref{alg:sliptrap}, 
\begin{align}
-\aN_{im}\V D_{m}^{F}= & -\frac{1}{\delta}\ \aN_{im}\left\{ \K_{pm}\left(\V Q^{+}\right)-\K_{pm}\left(\V Q^{-}\right)\right\} \slipW_{p}^{F}\nonumber \\
= & -\aN_{im}\left\{ \partial_{k}\K_{pm}\right\} \ \Rot_{kj}\Delta\V Q_{k}^{RFD}\slipW_{p}^{F}+\mathcal{R}\left(\delta^{2},\delta^{2}\right)\nonumber \\
= & -\N_{im}\left\{ \partial_{k}\K_{pm}\right\} \ \Rot_{kj}\N_{jl}\K_{lq}\Mob^{qs}\slipW_{s}^{D}\slipW_{p}^{F}+\mathcal{R}\left(\D t,\D t^{1/2}\right)+\mathcal{R}\left(\delta^{2},\delta^{2}\right),\label{eq:DFexpand}
\end{align}
where we used (\ref{eq:DeltaQref}) in (\ref{eq:DFexpand}). In expectation,
we get the drift term 
\begin{align}
\av{-\aN_{im}\V D_{m}^{F}}= & -\N_{im}\left\{ \partial_{k}\K_{pm}\right\} \av{\slipW_{s}^{D}\slipW_{p}^{F}}\Mob^{sq}\K_{ql}\N_{lj}\Rot_{kj}+O\left(\delta^{2}\right)+O\left(\D t\right)\nonumber \\
= & -k_{B}T\ \N_{im}\left\{ \partial_{k}\K_{pm}\right\} \Mob^{pq}\K_{ql}\N_{lj}\Rot_{kj}+O\left(\delta^{2}\right)+O\left(\D t\right),\label{eq:DFfin}
\end{align}
where we have used that $\av{\slipW_{s}^{D}\slipW_{p}^{F}}=k_{B}T\delta_{sp}$.

Similarly, using the definition of $\V D_{r}^{S}$ from step \ref{step:DFDS2}
of Algorithm \ref{alg:sliptrap}, 
\begin{align}
\aN_{im}\aK_{pm}\aMob^{pr}\V D_{r}^{S}= & \frac{1}{\delta}\ \aN_{im}\aK_{pm}\aMob^{pr}\left\{ \Mob_{rt}\left(\V Q^{+}\right)-\Mob_{rt}\left(\V Q^{-}\right)\right\} \slipW_{t}^{F}\label{eq:DSapprox}\\
= & \N_{im}\K_{pm}\Mob^{pr}\left\{ \partial_{k}\Mob_{rt}\right\} \ \Rot_{kj}\Delta\V Q_{k}^{RFD}\slipW_{t}^{F}+\mathcal{R}\left(\D t,\D t^{1/2}\right)+\mathcal{R}\left(\delta^{2},\delta^{2}\right).\nonumber 
\end{align}
Hence, taking the mean of (\ref{eq:DSapprox}), gives 
\begin{align}
\av{\aN_{im}\aK_{pm}\aMob^{pr}\V D_{r}^{S}}= & \N_{im}\K_{pm}\Mob^{pr}\left\{ \partial_{k}\Mob_{rt}\right\} \Rot_{kj}\av{\Delta\V Q_{k}^{RFD}\slipW_{t}^{F}}+O\left(\D t\right)+O(\delta^{2})\nonumber \\
= & \N_{im}\K_{pm}\Mob^{pr}\left\{ \partial_{k}\Mob_{rt}\right\} \av{\slipW_{s}^{D}\slipW_{t}^{F}}\Mob^{sq}\K_{ql}\N_{lj}\Rot_{kj}+O\left(\D t\right)+O(\delta^{2})\nonumber \\
= & k_{B}T\ \N_{im}\K_{pm}\Mob^{pr}\left\{ \partial_{k}\Mob_{rs}\right\} \Mob^{sq}\K_{ql}\N_{lj}\Rot_{kj}+O\left(\D t\right)+O(\delta^{2}).\label{eq:DSfin}
\end{align}
Combining equations (\ref{eq:DNindex}), (\ref{eq:DSfin}) and (\ref{eq:DFfin}),
we may write 
\begin{eqnarray}
\av{-\aN_{im}\V D_{m}^{F}+\aN_{im}\aK_{pm}\aMob^{pr}\V D_{r}^{S}} & = & k_{B}T\ \left(\left\{ \partial_{k}\N_{ij}\right\} \Rot_{kj}+\N_{im}\K_{pm}\Mob^{pr}\left\{ \partial_{k}\K_{rl}\right\} \N_{lj}\Rot_{kj}\right)\nonumber \\
 &  & +O\left(\D t\right)+O(\delta^{2}).
\end{eqnarray}
Combining this with (\ref{eq:slipUUfinal}) and using equation (\ref{eq:DNindex})
gives the desired result (\ref{eq:EM_RFD_U}).

%\bibliographystyle{unsrt}
%\bibliography{12_home_donev_Papers_Papers_References}

\end{document}